\documentclass[useAMS,usenatbib]{mnras}
\pdfoutput=1

\usepackage{amsmath}
\usepackage{txfonts}
\usepackage{graphicx}
\usepackage{url}
\usepackage{color}

\newcommand{\ot}{}

\newcommand{\pd}{\partial}
\newcommand{\ud}{\ensuremath{\mathrm{d}}}

\newcommand{\altaffilmark}[1]{$^{ #1}$}

\title[Assessment of SN Turbulence Models]{A Critical Assessment of Turbulence Models for 1D Core-Collapse Supernova Simulations}

\author[B.~M\"uller]{Bernhard~M\"uller\altaffilmark{1}\thanks{E-mail: bernhard.mueller@monash.edu}
\\
\altaffilmark{1}{Monash Centre for Astrophysics, School of
  Physics and Astronomy, Monash University, Victoria
  3800, Australia}
\\
\href{mailto:bernhard.mueller@monash.edu}{bernhard.mueller@monash.edu}
}

\begin{document}

\label{firstpage}
\pagerange{\pageref{firstpage}--\pageref{lastpage}}
\pagerange{\pageref{firstpage}--23}

\maketitle

\begin{abstract}
It has recently been proposed that global or local turbulence models
can be used to simulate core-collapse supernova explosions in
spherical symmetry (1D) more consistently than with traditional approaches
for parameterised 1D models. However, a closer
analysis of the proposed schemes reveals important consistency
problems. Most notably, they systematically violate energy
conservation as they do not balance buoyant energy generation with
terms that reduce potential energy, thus failing to account for the
physical source of energy that buoyant convection feeds on. We also
point out other non-trivial consistency requirements for
viable turbulence models. The model of
\citet{kuhfuss_86} proves more consistent than the newly proposed
approaches for supernovae,
but still cannot account naturally
for all the relevant physics
for predicting explosion properties.
We perform
numerical simulations for a $20 M_\odot$ progenitor to further illustrate problems of 1D turbulence
models. If the buoyant driving term is formulated in a conservative
manner, the explosion energy of $\mathord{\sim}2\times 10^{51}\, \mathrm{erg}$ for the corresponding non-conservative turbulence
model is reduced to $<10^{48} \,\mathrm{erg}$ even though the shock expands continuously. This demonstrates that the conservation problem cannot be ignored. 
Although plausible energies can be reached using
an energy-conserving model when turbulent viscosity is included, it is doubtful whether the energy budget of the explosion is regulated by the same mechanism as in multi-dimensional models. We conclude that 1D turbulence models based on a spherical Reynolds decomposition cannot provide a more consistent approach to supernova explosion and remnant properties than other phenomenological approaches before some fundamental problems are addressed.
\end{abstract}


\section{Introduction}
\label{sec:intro}
According to our present understanding, the explosions of massive
stars as core-collapse supernovae depend critically on the breaking of
spherical symmetry in the supernova core
\citep{janka_12,foglizzo_15,mueller_16c} except in the case of the
least massive progenitor stars \citep{kitaura_06}. In the
neutrino-driven paradigm, the breaking of spherical symmetry is
mediated by two instabilities, namely buoyancy-driven convection
\citep{herant_94,burrows_95,janka_95} and the standing-accretion shock
instability \citep{blondin_03,foglizzo_07}, which manifests itself in
the form of global sloshing or spiral motions of the shock.  The
resultant multi-dimensional (multi-D) fluid flow aids neutrino heating through a
variety of interrelated effects, e.g.\ by mixing hot neutrino-heated
and colder material from the vicinity of the shock, by providing
turbulent pressure \citep{burrows_95,murphy_12}, by providing
heating close to the shock by secondary shocks \citep{mueller_12b}, and
by turbulent dissipation \citep{mabanta_18}.

There have been attempts to distil these effects back into an
effective one-dimensional (1D) description using
an appropriate turbulence model. On the one hand, such a 1D
turbulence model for the supernova core may lead to a better
conceptual understanding of the role of multi-D effects
\citep{murphy_11,mabanta_18}.  On the other hand, one might hope that
effective 1D models of neutrino-driven supernovae could provide an
efficient way to predict the ``explodability'' and even the explosion
properties across a population of progenitor models at a cheaper cost
than full-blown multi-D models, but with greater rigour and
consistency than more parameterised approaches like those
of
\citet{oconnor_10,ugliano_12,perego_15,sukhbold_16,mueller_16a}.

The simplest approach of adopting the mixing-length theory (MLT) for
stellar convection \citep{biermann_32,boehm_58} to the supernova
problem already dates back to the 1980s \citep{mayle_85,wilson_88}. 
However, the extra convective energy transport provided by convection within the framework
of MLT alone does not improve the heating
conditions to such a degree as to allow explosions in spherical
symmetry \citep{huedepohl_phd,mirizzi_16}.

More general turbulence models have been proposed to capture
multi-D effects in the heating region more adequately
\citep{murphy_12,mabanta_18}.
Only recently have there been efforts to use such
turbulence models predictively. \citet{mabanta_18} 
incorporated a 1D turbulence model into steady-state solutions for the
accretion flow onto a proto-neutron star to derive the reduction of the
critical neutrino luminosity for shock revival \citep{burrows_93} due to
multi-D effects. Following up on their
earlier work, \citet{mabanta_19} went on
to include a simple turbulence model in dynamical
simulations with a view to studying the explodability
of supernova progenitors. Using a light-bulb
model for neutrino heating and cooling, they found
that their turbulence model roughly reproduces
the reduction of the critical luminosity
in multi-D for a reasonable choice of model parameters.
Shortly thereafter, \citet{couch_19}
presented a time-dependent 1D turbulence model, which they
coupled to the neutrino transport code of
\citet{oconnor_18b} and then used to explore the
systematics of explodability, and explosion and
remnant properties across the stellar mass range.
One might thus hope that such effective 1D turbulence models
can furnish a more  ``consistent'' approach
to the progenitor-explosion connection
than  current phenomenological models.

In this paper, we critically examine this idea.
We shall argue that consistency and crucial physical principles such as energy conservation are
difficult to achieve, especially during the explosion 
phase. 
We show that even the time-dependent approach of
\citet{couch_19} still suffers from inconsistencies.
To remedy these, one can draw on the extensive literature on
generalisations of mixing-length theory, which have
been studied since the 1960s
\citep[e.g.][]{spiegel_63,unno_67,eggleton_83,kuhfuss_86,gehmeyr_91,canuto_93,canuto_98,wuchterl_98}.
Classic time-dependent convection models
as developed by \citet{kuhfuss_86} and \citet{wuchterl_98}
offer solutions to most of the inconsistencies
in the models of \citet{mabanta_19}
and \citet{couch_19},
but even then  the behaviour of 1D turbulence models
during the explosion phase is not fully satisfactory. 
We illustrate the remaining problems for
a $20 M_\odot$ progenitor by implementing the 
model of \citet{kuhfuss_86} in the  neutrino hydrodynamics code \textsc{CoCoNuT} \citep{mueller_10,mueller_15a} with some necessary
adaptations for the core-collapse supernova problem.
The goal of our analysis and our numerical experiments is primarily
to illustrate the pitfalls that crop up when 
one seeks to model supernova explosions in 1D by including the effects of 
turbulence. We do not aim to present a consistent solution 
to all of these problems, which does not appear within reach at the 
moment.

This  paper is organised as follows:
In Section~\ref{sec:background} we discuss
recently proposed 1D turbulence models
and analyse to what extent they meet
important physical consistency criteria.
In Section~\ref{sec:minimal_model}, we
discuss the 1D turbulence model
of \citet{kuhfuss_86} and show that it
meets essential physical consistency criteria
such as total energy conservation and
compatibility with the second law of
thermodynamics. We also describe its
implementation in the
hydrodynamics code \textsc{CoCoNuT}.
We then present results for a $20 M_\odot$
star as test case using several variations
of the turbulence model of 
\citet{kuhfuss_86} in
Section~\ref{sec:results} and conclude by discussing future
perspectives for 1D explosion modelling
in Section~\ref{sec:conclusions}.

\section{Recent 1D Turbulence Models for Core-Collapse Supernovae}
\label{sec:background}
Both \citet{mabanta_19}
and \citet{couch_19} start from a spherical Reynolds decomposition 
of the fluid equations, from which they discard some higher-order terms
before they apply closure relations. Specifically,
they ignore the turbulent mass flux in the spirit of the anelastic
approximation, so that no turbulent correlation terms appear in the continuity equation. The common starting point
for both models can be written as
\begin{eqnarray}
\label{eq:gl1}
\pd_t \hat{\rho} + \nabla \cdot \langle \hat{\rho} \hat{\mathbf{v}}\rangle&=&0, \\
\label{eq:gl2}
\pd_t \langle \rho \mathbf{v}\rangle
+\nabla\cdot (\hat{\rho} \hat{\mathbf{v}}\ot \hat{\mathbf{v}})
+\nabla \hat{P} &=& \hat{\rho} \mathbf{g}
-\nabla \cdot \langle \rho \mathbf{R}\rangle, \\
\pd_t \langle \rho e \rangle
+\nabla \cdot [(\hat{\rho}\hat{e}+\hat{P})\mathbf{\hat{v}}]
&=&\hat{\rho} \mathbf{\hat{v}}\cdot \mathbf{g}
-\nabla \cdot (\mathbf{\hat{v}} \cdot \langle \rho \mathbf{R}\rangle )
\nonumber
\\
\label{eq:gl3}
&&-\nabla \cdot \mathbf{F}_\mathrm{conv} + \langle \rho ' \mathbf{v}'\rangle \cdot
\mathbf{g} +
\dot{\hat{q}}_\nu.
\end{eqnarray}
Here $\rho$, ${v_i}$, $P$, and $e$ denote the fluid
density, velocity, pressure, and total (internal+kinetic) energy density,
 $\mathbf{g}$ denotes the gravitational acceleration,
and $\dot{q}_\nu$ is the volumetric neutrino heating rate.
$\mathbf{R}$ and $\mathbf{F}_\mathrm{conv}$ are the
Reynolds stress tensor and the turbulent energy flux obtained
from the Reynolds decomposition. Spherical Reynolds averages
are denoted by angled brackets, or by carets for individual variables.
Primes, as used in the term $ \rho ' {v_i}'$ for buoyant
energy generation, denote fluctuating quantities.

\subsection{Model of \citet{mabanta_19}}
\label{sec:mabanta}
\citet{mabanta_19} impose a closure relation by essentially
assuming that the magnitude of turbulent radial velocity fluctuations $\delta v$ is constant
within the gain region, and that 
turbulent dissipation $\varepsilon$ balances buoyant energy generation 
if integrated over the entire gain region. With 
buoyant driving parameterised as $\beta \dot{Q}_\nu$ 
in terms of the volume-integrated heating rate $\dot{Q}_\nu$
and a calibrated parameter $\beta$, and turbulent
dissipation scaling as $\varepsilon\approx  \delta v^3/\Lambda $ in terms
of the typical value
of velocity perturbations $\delta v$
(which can formally defined as a root-mean-square
average  $\delta v=\langle v'^2\rangle^{1/2}$)
and an effective dissipation length $\Lambda$ (taken to
be the width of the gain region), one obtains
\begin{equation}
\delta v=\sqrt[3]{\frac{\beta \dot{Q}_\nu \Lambda}{M_\mathrm{gain}}},
\end{equation}
as also derived by \citet{mueller_15a}. The Reynolds
stress tensor is assumed to be diagonal with equipartition
between the radial and non-radial components so that
$R_{rr}=\hat{\rho} \,\delta v^2$ and
$R_{\theta\theta}=R_{\varphi\varphi}=\hat{\rho}\, \delta v^2/2$.
The turbulent dissipation $\varepsilon \sim \delta v^3/\Lambda $ 
is also used to supply the source term
$\langle \rho ' {v_i}'\rangle \mathbf{g}$
for buoyant energy generation in Equation~(\ref{eq:gl3}).

To obtain the convective flux, \citet{mabanta_19}
use a fit for the convective luminosity
$L_\mathrm{conv}=4\pi r^2 F_\mathrm{conv}$
that is inspired by an analysis of multi-D simulations,
\begin{equation}
\label{eq:lconv}
L_\mathrm{conv}=\alpha \dot{Q}_\nu \tanh\frac{r-r_\mathrm{g}}{h},
\end{equation}
Here $r$ is the radial coordinate, $r_\mathrm{g}$ is
the gain radius, $h$ is an appropriately chosen transition
width, and $\alpha$ is a calibrated dimensionless parameter.
\citet{mabanta_18} use Equation~(\ref{eq:lconv})
only up to the shock, where $L_\mathrm{conv}$ plummets to zero.
The term 
$\nabla \cdot (\mathbf{\hat{v}} \cdot \langle \rho \mathbf{R}\rangle )$
for the work exerted by Reynolds stresses is neglected in
the energy equation.

Although the prescriptions for the Reynolds stresses
and the convective luminosity  are in line with multi-D simulations,
there is an obvious question about energy conservation in this model.
Integrating Equation~(\ref{eq:gl3}) under the assumption
that turbulent dissipation locally balances buoyant energy generation
results in
\begin{eqnarray}
    \frac{\pd }{\pd t}
    \int 
    \rho e
    \,\ud V &=&
    \int \hat{\rho} \mathbf{\hat{v}}\cdot\mathbf{g} \,\ud V
    +\int  
    \rho \varepsilon \, \ud V
    +\int \dot{q}_\nu \, \ud V
    \\
    &=&\int   \hat{\rho} \mathbf{\hat{v}}\cdot\mathbf{g} \,\ud V
    +(1+\beta) \dot{Q}_\nu.
    \nonumber
\end{eqnarray}
The work done by gravitational forces on the spherically
averaged flow does not violate energy conservation and cancels
if gravitational potential energy is included in the budget
\citep{shu}. Similarly, the contribution  $\dot{Q}_\nu$ of neutrino heating
does not violate total energy conservation if it is obtained
from a conservative neutrino transport scheme. However,
the turbulence model introduces an extra source term
 $\beta \dot{Q}_\nu$ that violates energy conservation.
As we shall see, such a source term also appears in the
model of \citet{couch_19}.

It has been argued (Q.~Mabanta, private communication;
\citealp{couch_19}) that energy is still effectively
conserved if one accounts for the available free energy 
associated with convectively unstable gradients. Although this
argument is not entirely incorrect, it does not convincingly
justify the use of a 1D turbulence model that does not manifestly
conserve energy, but rather points to a loophole in the model:
In the full multi-D problem, energy is always \emph{strictly} conserved, so the reservoir of available free energy 
associated with unstable gradients must be \emph{depleted} as turbulent kinetic energy grows. Convective motions feed on
$P \, \ud V$ work of the expanding and contracting
bubbles analogous to a heat engine \citep{herant_94,mueller_15a},
and on the gravitational potential energy of the bubbles, {i.e.,
the reservoir of free energy associated with unstable
gradients forms a part of the reservoir of
internal and potential energy.}
Any increase in turbulent kinetic energy must 
therefore be balanced
by a decrease in internal or potential energy. To ensure
total energy conservation, a 1D turbulence model needs to
reflect this reshuffling of internal and potential energy by
consistently including all the necessary turbulent correlation terms. If there is a source term for the turbulent kinetic
energy, there must be a corresponding sink term
for internal or potential energy to account for the
fact that the available reservoir of free energy associated
with unstable gradients is drained as convective motions
develop.

But on the other hand, the source term
$\langle \rho ' \mathbf{v}'\rangle \cdot \mathbf{g}$
in the Reynolds-averaged energy equation~(\ref{eq:gl3})
is formally correct, so how is it possible that
the multi-D hydro equations conserve total energy
whereas the spherically Reynolds-averaged equation do not?
The answer is that the system 
(\ref{eq:gl1}-\ref{eq:gl3}) does \emph{not} include all terms
from the Reynolds decomposition
consistently: Specifically, it ignores
the turbulent mass flux $\langle \rho' \mathbf{v}'\rangle$
in Equation~(\ref{eq:gl1}) and thus neglects the (instantaneous)
reduction of potential energy due to changes in the density
profile, but includes it in the source
term in Equation~(\ref{eq:gl3}).
In Section~\ref{sec:minimal_model}, we shall further discuss to what extent this problem can be fixed.

There is another potential issue with the model of
\citet{mabanta_19}.
If Equation~(\ref{eq:lconv}) for the convective luminosity
is only used up to the shock, then this
implies that the term 
$\nabla \cdot \mathbf{F}_\mathrm{conv}$
in the energy equation will have a delta-function
spike at the shock, and it is unclear whether
such a feature can be handled properly by the
hydrodynamics solver, i.e.\ whether it affects
the propagation speed of the shock in an unphysical manner.

 \subsection{Model of \citet{couch_19}}
 \label{sec:couch}
 \citet{couch_19} solve a separate evolution
 equation for the turbulent velocity fluctuations 
 $\delta v$ and
 compute the turbulent flux in the energy equation
 using an MLT closure. They assume the
 same closure 
 $R_{rr}=2 R_\mathrm{\theta\theta}=
 2 R_{\varphi\varphi}=\hat{\rho} \,\delta v^2$
 for
 the Reynolds stress tensor as \citet{mabanta_19}.
 In their model, the momentum
 and energy equations come down to
 \begin{eqnarray}
     \frac{\pd \hat{\rho} \hat{v}_r}{\pd t}
    +\frac{1}{r^2}\frac{\pd}{\pd r}
    \left[
    r^2(\hat{\rho} \hat{v}_r^2+\hat{P}+ \hat{\rho}\, \delta v^2)
    \right]
    &=&-\hat{\rho} g\\
    \frac{\pd \hat{\rho} \hat{e}}{\pd t}
    +\frac{1}{r^2}\frac{\pd}{\pd r}
    \left[r^2 v_r (\hat{\rho} \hat{e}+\hat{P}+\hat{\rho}\, \delta v^2)
    -r^2 \hat{\rho} D \nabla \hat{e}
    \right]
    &=&-\hat{\rho} \hat{v}_r g+\hat{\rho} \frac{\delta v^3}{\Lambda} \nonumber\\
\label{eq:gl3_couch}    
    &&+\dot{\hat{q}}_\nu,
 \end{eqnarray}
where we have neglected the neutrino momentum source
term, which is irrelevant for our discussion.
Here $D= \delta v\, \Lambda/3$ is an MLT diffusion coefficient, and the mixing length $\Lambda$ is chosen
as a multiple
of the pressure scale height
$\Lambda=\alpha \hat{P}/(\hat{\rho} g)$, where
$\alpha$ is a tunable parameter of
order unity. Note that we have tacitly
corrected a typo in their
Equation~(26), where the term
$\hat{\rho} D \nabla e$  should not be multiplied by the
radial velocity $\hat{v}_r$. There is likely another typo
in their Equation~(25), which should contain
a fictitious force term
$\hat{\rho}({R}_{\theta\theta}+{R}_{\varphi\varphi})/r
=\hat{\rho}\, \delta v^2/r$
on the right-hand side (RHS). Moreover, the energy equation
appears to assume an isotropic Reynolds stress
tensor, and is inconsistent with the assumed
form of $\mathbf{\hat{R}}$. However, these problems
are somewhat peripheral to our further discussion.

For the evolution of the turbulent kinetic energy,
\citet{couch_19} propose the following
equation:
\begin{equation}
\label{eq:gl4_couch}
    \frac{\pd \hat{\rho}\, \delta v^2}{\pd t}
    +\frac{1}{r^2}\frac{\pd}{\pd r}
    \left[r^2 (\hat{\rho} \hat{v}_r \delta v^2
    -\hat{\rho} D \nabla \delta v^2)
    \right]
    =
    \hat{\rho}\, \delta v\, \omega^2_\mathrm{BV} \Lambda
    -\hat{\rho}\frac{\delta v^3}{\Lambda}
    .
\end{equation}
Here, the turbulent dissipation term $\delta v^3/\Lambda$
appears as a local sink term, and the source
term is expressed in terms of the Brunt-V\"ais\"al\"a
frequency $\omega_\mathrm{BV}$
in a manner consistent with
classical, time-dependent MLT
(cp.\ Section~2.1 in \citealt{mueller_16c}).

Including a separate time-dependent equation for the
turbulent kinetic energy makes the reshuffling
of energy between the spherically-averaged bulk flow
and the turbulent fluctuations more transparent,
but still does not solve the problem of
energy conservation, and even introduces
further ambiguities.
From the form of the
equations and the discussion in \citet{couch_19} it appears
that $\hat{\rho}\hat{e}$ in the
energy equation~(\ref{eq:gl3_couch}) is  \emph{not}
meant to include the turbulent kinetic energy, 
so that one needs to add Equations~(\ref{eq:gl3_couch})
and (\ref{eq:gl4_couch}) to analyse total energy conservation.
As in Section~\ref{sec:mabanta}, this again leads to an additional source term
$\int \hat{\rho}\, \delta v \, \omega_\mathrm{BV}^2 \Lambda \,\ud V$
that breaks total energy conservation.
Since the typical turbulent velocity must be related
to the amount of neutrino heating just as in multi-D
in order to maintain a stationary, slightly
superadiabatic gradient, the
rate of artificial energy injection by
buoyant driving is also
of the same scale as before, i.e.\ roughly of
order of the volume-integrated neutrino heating rate
$\dot{Q}_\nu$ itself.

Moreover, the model leaves us with a consistency problem.
While it is implicitly assumed that 
$\hat{\rho} \hat{e}$ does \emph{not} include the turbulent kinetic
energy -- which is why the dissipation term
appears as a source in Equation~(\ref{eq:gl3_couch}) --
the term for the work done by Reynolds stresses
is written as a divergence
 in Equation~(\ref{eq:gl3}). In Equation~(\ref{eq:gl3}),
however, $\langle \rho e \rangle$ implicitly \emph{includes}
the turbulent kinetic energy as correctly
recognised in \citet{mabanta_19}. If one 
subtracts the turbulent kinetic energy equation
from the total energy equation, one
must instead treat Reynolds stresses as a body
force so that their contribution on the RHS of
the energy equation is
$-\mathbf{\hat{v}} \cdot (\nabla\cdot \hat{\rho} \mathbf{R})$
instead of
$-\nabla\cdot (\mathbf{\hat{v}} \cdot \rho \mathbf{R})$
(see also below in Section~\ref{sec:minimal_model}).

This may seem a trivial issue, but it is enlightening
to analyse its potential repercussions. Assuming for
the sake of simplicity that the turbulent pressure
$P_\mathrm{t}$
is actually isotropic, then including the
work of Reynolds stresses as a divergence as in
Equation~(\ref{eq:gl3_couch}), implies that
the evolution equation for the internal energy
density $\hat{\epsilon}$
of the spherically symmetric bulk flow
is changed to 
\begin{equation}
\frac{\ud \hat{\epsilon}}{\ud t}
=-(\hat{P}+P_\mathrm{t}) \frac{\ud (1/\hat{\rho})}{\ud t},
\end{equation}
in the absence of heating and cooling terms
from neutrinos and dissipation, instead
of 
\begin{equation}
\frac{\ud \hat{\epsilon}}{\ud t}
=-\hat{P} \frac{\ud (1/\hat{\rho})}{\ud t}.
\end{equation}
This is tantamount to an artificial entropy
source term
\begin{equation}
\hat{T}\frac{\ud \hat{s}}{\ud t}
=-P_\mathrm{t} \frac{\ud (1/\hat{\rho})}{\ud t}.
\end{equation}
It is noteworthy that this term can become
negative, i.e.\ the turbulence model implicitly allows
for a decrease of entropy even in the absence of physical
source terms for cooling.

There is yet another problem with
the model of \citet{couch_19} that concerns the convective energy flux
$\mathbf{F}_\mathrm{conv}$.
The model essentially assumes that $\mathbf{F}_\mathrm{conv}$
can be computed by extrapolating the total energy density
$e$ to the original position of the convective bubbles
using the local gradient to obtain the fluctuating part $e'$,
\begin{equation}
\label{eq:wrong_eprime}
   e'=\left(\hat{\epsilon+}\frac{\hat{v}_r^2}{2}\right)'=
   \Lambda \frac{\pd \hat{e}}{\pd r}
   =\Lambda \frac{\pd \hat{\epsilon}}{\pd r}+\Lambda \frac{\pd \hat{v}_r^2/2}{\pd r}.
\end{equation}
This, however, leads to unphysical results. Let us first
consider the fluctuations of the internal energy density
$\hat{\epsilon}$. To obtain the correct MLT flux, one needs
to account for the $P\, \mathrm{d}V$ done
by the convective bubbles as they contract and expand
while adjusting to the ambient pressure
\citep{huedepohl_phd,mirizzi_16}. If the expansion/contraction
is adiabatic, one obtains
\begin{equation}
\label{eq:eprime}
    e'=\Lambda \left[\frac{\pd \hat{\epsilon}}{\pd r}+
    \hat{P} \frac{\pd}{\pd r}\left(\frac{1}{\hat{\rho}}\right)\right].
\end{equation}
By expressing $\pd \epsilon/\pd r$ in terms of
the entropy and density gradients using the first
law of thermodynamics, one obtains
an important corollary: \emph{If the entropy gradient
vanishes, then the convective energy flux also vanishes}
(assuming that there are no composition gradients).
If one uses Equation~(\ref{eq:wrong_eprime}) this is no longer
guaranteed.

There is also a concern about the turbulent transport of bulk kinetic
energy:
This effect is included in Equation~(\ref{eq:gl3_couch})
via Equation~(\ref{eq:wrong_eprime}), but there is no corresponding term
for the turbulent transport of momentum (i.e.\ turbulent
viscosity) in Equation~(\ref{eq:gl3_couch}).

\section{The Energy-Conserving Turbulence Model of Kuhfuss (1986)}
\label{sec:minimal_model}
\subsection{Description and Discussion of Individual
Terms}
\label{sec:kuhfuss1}
Most of the problems discussed in Sections~\ref{sec:mabanta}
and \ref{sec:couch} are in fact nicely solved by 
time-dependent one-equation turbulence models that have been developed 
for stellar evolution \citep{stellingwerf_82,kuhfuss_86,wuchterl_98},
originally motivated by the problem of pulsations of RR Lyrae stars
and Cepheids.
Here we shall use the work of \citet{kuhfuss_86} (with
some modifications by \citealt{wuchterl_98})
as a starting point. In their approach, the momentum equation
in conservation form can be written as
\begin{equation}
\frac{\pd \hat{\rho} \hat{v}_r}{\pd t}
+\frac{1}{r^2}\frac{\pd r^2 \hat{\rho} \hat{v}_r}{\pd r}
+\frac{\pd (\hat{P}+P_\mathrm{t})}{\pd r}
=-\hat{\rho} g+
\frac{4}{3r^3}\frac{\pd }{\pd r}
\left[r^3 \mu
\left(\frac{\pd \hat{v}_r}{\pd r}
-\frac{\hat{v}_r}{r}\right)\right].
\end{equation}
The underlying assumption about the form of the Reynolds stress
tensor differs slightly from \citet{mabanta_19}
and \citet{couch_19}; it is decomposed into a trace component --
the turbulent pressure $P_t$ -- and a trace-free component modelled
after the viscous Navier-Stokes equations. This trace-free term
gives rise to the additional turbulent viscosity term on the
RHS with a turbulent dynamic viscosity\footnote{Note
that our notation is different from
\citet{kuhfuss_86}, where
$\mu$ is used for the \emph{kinematic}
turbulent viscosity.} $\mu$,
which is expressed in the spirit of MLT as
$\alpha_\mu \hat{\rho} \Lambda\, \delta v $,
where $\alpha_\mu$ is a dimensionless coefficient of order unity. Although the assumed form of the trace-free term can be
criticised as \emph{ad hoc}, it has the virtue of ensuring that
it can be matched with a viscous term in the energy equation
that can be expressed as a flux divergence (to ensure energy conservation)
and always results in an increase of fluid entropy.

\citet{kuhfuss_86} and 
\citet{wuchterl_98}
also formulate an evolution equation
for the turbulent kinetic energy $\varpi$. 
With one important modification, the equation for $\varpi$ can be written as
\begin{multline}
\label{eq:kuhfuss_varpi}
    \frac{\pd \hat{\rho} \varpi}{\pd t}
    +
    \frac{1}{r^2}\frac{\pd}{\pd r}
    \left[r^2 \hat{v}_r \hat{\rho} \varpi\right]
    +\frac{P_t}{r^2}\frac{\pd r^2 \hat{v}_r}{\pd r}\\
    +\frac{\alpha_\varpi}{r^2}\frac{\pd}{\pd r}
    \left(r^2 \hat{\rho} \Lambda \varpi^{1/2} \frac{\pd \varpi}{\pd r}\right)
    =-\langle \rho' v'\rangle  g
    -C_\mathrm{D} \, \hat{\rho} \frac{\varpi^{3/2}}{\Lambda},
\end{multline}
in Eulerian form.
Here, $\alpha_\varpi$ and $C_\mathrm{D}$
are dimensionless coefficients for the
turbulent diffusion of turbulent kinetic energy and turbulent dissipation.
Different from 
\citet{kuhfuss_86} and \citet{wuchterl_98},
we have
omitted the  viscous dissipation term
from the turbulent kinetic energy equation.
Although it seems plausible that
the turbulent viscosity should
should initially feed kinetic energy from
the bulk flow into disordered small-scale
motion (i.e.\ into turbulent kinetic energy),
this has undesirable consequences. The problem is
that the viscous heating term is \emph{linear} in $\delta v$ just like
the buoyant driving term.
If included in the equation for $\varpi$,
it would act like a destabilising gradient
wherever $\pd \hat{v}_r/\pd r -\hat{v}_r/r \neq 0$ 
(i.e.\ almost everywhere)
with  $|\pd \hat{v}_r/\pd r -\hat{v}_r/r|$ taking 
the place of the Brunt-V\"ais\"al\"a frequency
$\omega_\mathrm{BV}$ as the growth rate.
From the viewpoint of numerical stability,
such behaviour would be disastrous at the shock, 
but it is also clearly unphysical. Shifting
the viscous heating term to the internal energy 
equation seems the only viable solution.

Like Equation~(\ref{eq:gl4_couch}), the
turbulent energy equation includes terms for the advection
and the turbulent diffusion of turbulent kinetic energy (the two flux divergence term on the left-hand side). However, there is also a term that effectively
accounts for $P\, \ud V$ term on the ``eddy gas''. Including
this term in Equation~(\ref{eq:kuhfuss_varpi})
ensures that the work exerted by the turbulent pressure correctly
enters as $-\mathbf{\hat{v}}\cdot \nabla P_\mathrm{t}$
in the corresponding equation for $\hat{\rho} \hat{e}$ \emph{without} turbulent
kinetic energy, and hence does not change the internal energy density of
the bulk flow. Apart from dimensionless coefficients of order unity,
the source term for buoyant driving 
 and the dissipation term
are essentially the same as in Equation~(\ref{eq:gl4_couch});
the driving term can again be expressed in terms of
the Brunt-V\"ais\"al\"a frequency as
$-\langle \rho' v'\rangle  g 
\propto \hat{\rho}\, \delta v\, \omega_\mathrm{BV}^2  \Lambda $.

In the theory of \citet{kuhfuss_86}, total energy conservation is ensured by
consistently including the mass-specific turbulent kinetic energy $\varpi$
in the energy equation.
\citet{kuhfuss_86} originally formulated
an extended internal energy equation including $\varpi$ as the
internal energy of an ``eddy gas'', but this equation can be
easily recast into a total energy equation
analogous to Equation~(\ref{eq:gl3}),
\begin{multline}
\label{eq:kuhfuss_energy}
\frac{\pd \hat{\rho} (\hat{e}+\varpi)}{\pd t}
+\frac{1}{r^2}\frac{\pd  }{\pd r}
\left[
r^2 \hat{\rho} (\hat{e}+\varpi) v_r
+r^2 (\hat{P}+P_t) \hat{v}_r\right.\\
\left.
+r^2 (F_\mathrm{conv}+
F_\mathrm{\varpi}+F_\mathrm{visc})
 \right]
=-\hat{\rho} \hat{v}_r g .
\end{multline}
Here, $F_\mathrm{conv}$ and $F_\mathrm{\varpi}$ denote
the convective flux of internal energy and turbulent kinetic energy,
and $F_\mathrm{visc}$ is the energy flux from viscous turbulent stresses.
Equation~(\ref{eq:kuhfuss_energy}) is manifestly conservative because
all the turbulent effects are lumped into flux divergence terms.
In accordance
with Equation~(\ref{eq:eprime}), the convective
energy flux is calculated as
\begin{equation}
F_\mathrm{conv}
=\langle \rho v_r '\epsilon'\rangle
=
-\alpha_e \hat{\rho}\,\delta v\,
\Lambda \left[\frac{\pd \hat{\epsilon}}{\pd r}+\hat{P} \frac{\pd}{\pd r}\left(\frac{1}{\hat{\rho}}\right)\right],
\end{equation}
where $\alpha_e$ is another adjustable
dimensionless parameter for turbulent energy
transport. The viscous energy
flux is given by
\begin{equation}
   F_\mathrm{visc}
=
-\frac{4}{3}\mu 
\hat{v}_r
\left(\frac{\pd \hat{v}_r}{\pd r}
-\frac{\hat{v}_r}{r}\right),
\end{equation}
and as per Equation~(\ref{eq:kuhfuss_varpi}),
the convective flux of turbulent kinetic energy
is 
\begin{equation}
     F_\mathrm{\varpi}
     =
     -\alpha_\varpi
    \hat{\rho} \Lambda \, \delta v
    \frac{\pd \varpi}{\pd r}
\end{equation}

For further analysis, it is useful to consider the internal energy
equation for the spherical background flow as well, which 
reads
\begin{multline}
\label{eq:kuhfuss_epsilon}
    \frac{\pd \hat{\rho} \hat{\epsilon}}{\pd t}
    +
    \frac{1}{r^2}\frac{\pd}{\pd r}
    \left[r^2 {\hat{\rho}\hat{v}_r \hat{\epsilon}} \right]
    +\frac{\hat{P}}{r^2}\frac{\pd r^2 \hat{v}_r}{\pd r}
    -\frac{\alpha_e}{r^2}\frac{\pd}{\pd r}
    \left[r^2 \hat{\rho} \Lambda \varpi^{1/2} 
    \left(\frac{\pd \hat{\epsilon}}{\pd r}+\hat{P}
    \frac{\pd (1/\hat{\rho})}{\pd r}\right)\right]\\
    =\langle \rho' v'\rangle  g
    +C_\mathrm{D} \, \hat{\rho} \frac{\varpi^{3/2}}{\Lambda}+
        \frac{4}{3}\alpha_\mu \hat{\rho} \Lambda\, \delta v \left[\frac{\pd \hat{v}_r}{\pd r}-\frac{\hat{v}_r}{r}\right]^2,
\end{multline}
where $\alpha_e$ is another dimensionless coefficient of order unity and
Equation~(\ref{eq:eprime}) has been used to express
the convective flux of internal energy.
Note that the viscous dissipation term appears
in the internal energy equation for reasons
explained above.

Equations~(\ref{eq:kuhfuss_energy}), (\ref{eq:kuhfuss_varpi}), and (\ref{eq:kuhfuss_epsilon}) ensure
that i) entropy is conserved during expansion
and contraction in the absence of diffusive flux and source/sink term,
that ii) the convective flux transports energy into the direction
of negative entropy gradients, and that iii) total energy is conserved. Energy conservation is achieved by including a \emph{sink} term due to buoyant driving in the internal energy equation, and not including a source term for buoyant driving in the total energy equation.

\subsection{The Kuhfuss Model Interpreted in the Framework
of Favre Decomposition}
\label{sec:kuhfuss_favre}
It may appear as though the solution for energy conservation
in the model of \citet{kuhfuss_86} were somewhat 
\emph{ad hoc} and still inconsistent, because it merely relies on neglecting
the turbulent mass flux $\langle \rho' v_r'\rangle$ in different
equations than \citet{mabanta_19} and \citet{couch_19}.
But this appearance is deceptive.
The Kuhfuss model is in fact consistent if we re-interpret
the equations as arising from a Favre
decomposition of the flow based on mass-weighted
averages \citep{favre_65} for mass-specific
quantities (e.g., $\epsilon$, $\mathbf{v}$, 
$\varpi$, and mass fractions $X_i$)
instead of a Reynolds decomposition.
Using tildes to denote Favre averages
$\tilde{Y}=\langle \rho Y\rangle/\hat{\rho}$
for any variable $Y$ and double primes for fluctuations
around Favre averages, the 
equations for conservation of mass, momentum,
energy, partial masses, and for the turbulent
kinetic energy read,
\begin{gather}
\frac{\pd \hat{\rho} }{\pd t}
+
\nabla \cdot
(\hat{\rho}\tilde{\mathbf{v}}) =0,
\\
\frac{\pd \hat{\rho} \mathbf{v}}{\pd t} 
+
\nabla \cdot \hat{\rho} \tilde{\mathbf{v}} \ot \tilde{\mathbf{v}}
+
\nabla \cdot \hat{\rho} \tilde{\mathbf{R}}
+
\nabla \hat{P}
=
\hat{\rho} \mathbf{g},
\\
\frac{\pd }{\pd t}
\left(
\hat{\rho}\tilde{\epsilon}
+
\hat{\rho} \frac{|\tilde{\mathbf{v}}|^2}{2}
+
\hat{\rho} \varpi
\right)
+
\nabla \cdot
\Big [
(\hat{\rho}\tilde{\epsilon}+
\hat{\rho} \frac{|\tilde{\mathbf{v}}|^2}{2}
+\hat{P})
\tilde{\mathbf{v}}
+
\hat{\rho} \langle \epsilon'' \mathbf{v}''\rangle
\nonumber
\\
+
\langle P' \mathbf{v}''\rangle
+
\hat{\rho} \tilde{\mathbf{R}} \cdot \tilde{\mathbf{v}}
+
\hat{\rho} \tilde{\mathbf{v}} \varpi
+
\langle \rho \mathbf{v}'' \frac{|\mathbf{v}''|^2}{2}\rangle
\Big ]
=
\hat\rho
\tilde{\mathbf{v}}
\cdot \mathbf{g}.
\\
\frac{\pd \hat\rho \varpi}{\pd t}
    +\langle \mathbf{v}''\cdot \nabla P'\rangle
    +(\hat{\rho} \tilde{\mathbf{R}} \cdot \nabla) \tilde{\mathbf{v}}
    +\nabla \cdot (\hat{\rho} \tilde{\mathbf{v}} \varpi)
    +\langle \rho \mathbf{v}'' \frac{|\mathbf{v}''|^2}{2}\rangle
    =
    \langle \rho'\mathbf{v}''\rangle \cdot
    \mathbf{g}.
\\
\frac{\pd \hat{\rho } \tilde{X}_i}{\pd t} 
+
\nabla  \cdot
(\hat{\rho} \tilde{\mathbf{v}} \tilde{X}_i)
+\nabla
\cdot( \rho  \widetilde{\mathbf{v}'' X_i''})
=0.
\end{gather}
as derived in Appendix~\ref{sec:favre}.
Note that the turbulent kinetic energy is now
defined as the Favre average $\varpi=\widetilde{|\mathbf{v}''|^2}$/2.

The only terms that are not mirrored\footnote{
Bear in mind that the Reynolds stress tensor is broken
up into a trace part and a trace-free part
in the Kuhfuss model.} in the
model of \citet{kuhfuss_86} are the ones
containing turbulent fluctuations of the pressure
due to the assumption of instant horizontal
pressure equilibration. Different from the Reynolds-decomposed
equations, the total mass flux
$\langle\rho v_r \rangle$ factorises
as $\langle\rho v_r \rangle = \hat{\rho} \tilde{v}_r$.
This eliminates both the turbulent mass flux term
$\langle\rho' v'_r \rangle$ from the continuity
equation and the term $\langle\rho' v'_r g \rangle$
from the total energy equation, while  $\langle\rho' v'_r g \rangle$
still naturally appears as a source term in the kinetic
energy equation. Thus, the choice of discarding the 
$\langle\rho' v'_r g \rangle$ term in the energy equation
 is not actually an approximation in the model
 of \citet{kuhfuss_86}, and there is in fact no need
 to include extra terms in the Kuhfuss model to achieve
 full consistency with the spherically-averaged
 fluid equations. Instead, we simply need to treat
 some of the evolved quantities in the
 Kuhfuss model as Favre averages and solve
 the following equations for mass, momentum, internal energy, partial mass fractions $X_i$, and turbulent kinetic energy
$\varpi$:
\begin{equation}
\label{eq:kuhfuss_mass2}
\frac{\pd \hat{\rho}}{\pd t}
+\frac{1}{r^2}\frac{\pd r^2 \hat\rho \tilde{v}_r}{\pd r}
=0,
\end{equation}
\begin{equation}
\label{eq:kuhfuss_momentum2}
\frac{\pd \hat{\rho} \tilde{v}_r}{\pd t}
+\frac{1}{r^2}\frac{\pd r^2 \hat\rho \tilde{v}_r}{\pd r}
+\frac{\pd (\hat{P}+P_\mathrm{t})}{\pd r}
=-\hat{\rho} g+
\frac{4}{3r^3}\frac{\pd }{\pd r}
\left[r^3 \mu
\left(\frac{\pd \tilde{v}_r}{\pd r}
-\frac{\tilde{v}_r}{r}\right)\right],
\end{equation}
\begin{multline}
\label{eq:kuhfuss_energy2}
    \frac{\pd \hat\rho \tilde{\epsilon}}{\pd t}
    +
    \frac{1}{r^2}\frac{\pd}{\pd r}
    \left[r^2  \hat{\rho} \tilde{\epsilon} \tilde{v}_r\right]
    +\frac{\hat{P}}{r^2}\frac{\pd r^2 \tilde{v}_r}{\pd r}
    -\frac{\alpha_e}{r^2}\frac{\pd}{\pd r}
    \left[r^2 \hat{\rho} \Lambda \varpi^{1/2} 
    \left(\frac{\pd \tilde{\epsilon}}{\pd r}+\hat{P}
    \frac{\pd (1/\hat{\rho})}{\pd r}\right)\right]\\
    =\langle \rho' \tilde{v}_r''\rangle  g
    +C_\mathrm{D} \, \hat{\rho} \frac{\varpi^{3/2}}{\Lambda}+
        \frac{4}{3}\alpha_\mu \hat\rho \Lambda \tilde{v}_r'' \left(\frac{\pd \tilde{v}_r}{\pd r}-\frac{\tilde{v}}{r}\right)^2,
\end{multline}
\begin{equation}
\label{eq:kuhfuss_x2}
    \frac{\pd \hat\rho \tilde{X}_i}{\pd t}
    +
    \frac{1}{r^2}\frac{\pd}{\pd r}
    \left[r^2  \hat{\rho} \tilde{X}_i \tilde{v}_r\right]
    -\frac{\alpha_X}{r^2}\frac{\pd}{\pd r}
    \left[r^2 \hat{\rho} \Lambda \varpi^{1/2} 
    \left(\frac{\pd \tilde{X}_i}{\pd r}\right)\right]\\
    =0,
\end{equation}
\begin{multline}
\label{eq:kuhfuss_varpi2}
    \frac{\pd \hat{\rho} \varpi}{\pd t}
    +
    \frac{1}{r^2}\frac{\pd}{\pd r}
    \left[r^2 \tilde{v}_r \hat\rho \varpi\right]
    +\frac{P_t}{r^2}\frac{\pd r^2 \tilde{v}_r}{\pd r}\\
    -\frac{\alpha_\varpi}{r^2}\frac{\pd}{\pd r}
    \left(r^2 \hat\rho \Lambda \varpi^{1/2} 
    \frac{\pd \varpi}{\pd r}\right)
    =-\langle \rho' \tilde{v}_r''\rangle  g
    -C_\mathrm{D} \, \hat\rho \frac{\varpi^{3/2}}{\Lambda}.
\end{multline}
Here, the turbulent viscosity is defined analogously
to Section~\ref{sec:kuhfuss1} as
$\mu=\alpha_\mu \hat{\rho} \Lambda \, \delta v$,
and for reasons explained above, we still prefer to
feed the energy from turbulent viscous dissipation
directly into internal energy. 
Note that one can also formulate an equation
for the mole fractions $Y_i$ instead
of Equation~(\ref{eq:kuhfuss_x2}); its form is
completely analogous.

To close the system,  we assume that
the root-mean-square radial velocity fluctuations $\delta v$ are related to
the total turbulent kinetic energy as $\delta v=\varpi^{1/2}$,
reflecting the rough equipartition between radial and non-radial
velocity fluctuations in multi-D simulations.
Different from \citet{kuhfuss_86}, we shall use
$P_\mathrm{t} =\hat{\rho}\, \delta v^2 =\hat{\rho} \varpi$
(as originally proposed
by  \citealt{stellingwerf_82}) in our
numerical implementation, which makes for a larger force
from the turbulent pressure gradient
in the momentum equation. Although this is somewhat inconsistent, it does not fundamentally alter the structure and properties of the turbulence model. The source term $\langle \rho' v_r''\rangle$ can be expressed 
as $\langle \rho' v_r''\rangle=\delta \rho \, \delta v$ using
the MLT density contrast
\footnote{Note
that this is equivalent
to the expression $\hat{\rho}\, \delta v\, \omega_\mathrm{BV}^2 \Lambda$
ussed in \ref{sec:couch}.}
\begin{equation}
    \delta \rho =\Lambda \left(\frac{\pd \hat{\rho} }{\pd r}-\frac{1}{\tilde{c}_\mathrm{s}^2}\frac{\pd \hat{P}}{\pd r}\right).
\end{equation}
The model contains six adjustable non-dimensional parameters,
namely $\alpha$ as the ratio of the mixing length
to the pressure scale height, and the coefficient
 $C_\mathrm{D}$ for turbulent dissipation, $\alpha_\mu$
 for the turbulent viscosity, and $\alpha_e$, $\alpha_X$, and $\alpha_\varpi$ for the turbulent transport of bulk energy,
 mass fractions, and turbulent kinetic energy. In our implementation, we shall set all of these coefficient to
 unity
 ($C_\mathrm{D}=\alpha_\mu=\alpha_e=\alpha_\varpi=\alpha_X=1$) except for the mixing-length coefficient $\alpha$.
In fact, we shall see that $\alpha_e=\alpha_X=1$
must hold for consistency reasons given the
assumed form of the term $-\langle \rho' v_r''\rangle g$
for buoyant driving.

\subsection{Compatibility with the Second Law of
Thermodynamics}
\label{sec:entropy}
In the light of our discussion of entropy conservation
earlier in this paper,  the form of the
Equations~(\ref{eq:kuhfuss_mass2}-\ref{eq:kuhfuss_varpi2})
appears puzzling at first glance: It is easy to see
that the additional terms from turbulent stresses,
turbulent viscosity, and turbulent dissipation
either conserve or generate entropy and are
thus compatible with the second law of thermodynamics.
The sink term from buoyant driving in the internal
energy generation seems to violate the second law,
however.
It appears  that the price for energy conservation
is the possibility of a decrease in entropy in the bulk flow
due to the generation of turbulent kinetic energy, i.e.\
a violation of the second law of thermodynamics.
This is especially curious since we cannot blame this
sink term on the inconsistent treatment of the
turbulent mass flux anymore, which has simply disappeared
in the Favre decomposition.

The solution to this conundrum
is that this sink term cancels an extra entropy source
term that emerges from the convective energy flux term.
In order to see this, let us first
determine the change of entropy due to the convective
transport (denoted by the subscript ``conv'') 
of energy and mole fractions. It is convenient to consider
the convective derivative of  thermodynamic quantities,
whieh can be obtained from the fluxes in the conservation
equations in a straightforward manner, e.g.,
\begin{equation}
    \left(\frac{\ud \tilde\epsilon}{\ud t}\right)_\mathrm{conv}
    =-\frac{\nabla \cdot \mathbf{F}_\mathrm{conv}}{\hat{\rho}},
\end{equation}
for the internal energy. It is also advantageous not to
spell out the del operator in spherical polar coordinates.
With these considerations in mind, we obtain the rate of change
of the fluid entropy due to turbulent transport as
\begin{multline}
    \tilde{T} \left(\frac{\ud \tilde{s}}{\ud t}\right)_\mathrm{conv}
    =
    \left(\frac{\ud \tilde{\epsilon}}{\ud t}\right)_\mathrm{conv}-
    \sum_i \mu_i\left(\frac{\ud \tilde{Y_i}}{\ud t}\right)_\mathrm{conv}
    \\
    =
    \frac{1}{\hat{\rho}}
    \nabla \cdot \left\{
    \alpha_e \Lambda\, \delta v 
    \left[\hat{\rho} \nabla \epsilon
    +\hat{P} \nabla \left(\frac{1}{\hat{\rho}}\right)
    \right]
    \right\}
    -
    \sum_i \frac{\tilde{\mu}_i}{\hat{\rho} }
    \nabla \cdot
    \left(
    \alpha_X \hat{\rho} \Lambda\, \delta v \nabla \tilde{Y}_i
    \right),
\end{multline}    
where $\tilde{\mu}_i$ denotes the chemical potential for species
$i$ (and not the turbulent viscosity).\footnote{Note that in the strictest
sense the thermodynamic quantities
$\tilde{T}$,
$\tilde{\mu}_i$,
and $\hat{P}$ are not Favre/Reynolds averages,
but functions of the Favre/Reynolds averages
of $\hat{\rho}$, $\tilde{\epsilon}$,
and $\tilde{Y}_i$, i.e.\ they are
to be understood as
$\tilde{T}=T(\hat{\rho},\tilde{\varepsilon},\tilde{Y}_i)$, and so forth.
}
For this to be a conservation equation, we would need
the RHS to be of the form
$\hat{\rho}^{-1} \tilde{T} \nabla \cdot \mathbf{F}_s$, where $\mathbf{F}_s$
is an MLT  turbulent entropy flux proportional to
 $\Lambda \hat{\rho}  \, \delta v 
 \nabla \tilde{s}$. We can obtain
such a term by means of rearrangements
on the RHS if $\alpha_e=\alpha_X$ (which
we shall always assume henceforth), but
 are then  left with an extra term,
\begin{align}
\tilde{T} \left(\frac{\ud \tilde{s}}{\ud t}\right)_\mathrm{conv}
    &=
  \frac{1}{\hat{\rho}}
    \nabla \cdot \left\{
    \alpha_e \Lambda \, \delta v 
    \left[\hat{\rho} \nabla \epsilon
    +P \nabla \left(\frac{1}{\hat{\rho}}\right)
    -\sum \tilde{\mu}_i \hat{\rho}  \Lambda \, \delta v \nabla \tilde{Y}_i
    \right]
    \right\}
    \nonumber
    \\
    &
    \phantom{=}
    +
    \sum_i 
     \left(
    \alpha_e \Lambda \, \delta v \nabla \tilde{Y}_i
    \right)
    \nabla \tilde{\mu}_i
    \label{eq:dsdt_conv}
      \\
    &=
  \frac{1}{\hat{\rho}}
    \nabla \cdot \left(
    \alpha_e \Lambda \,  \hat{\rho} \, \delta v \,
    \tilde{T} \nabla  \tilde{s}
    \right)
    +
    \sum_i 
    \alpha_e \Lambda \, \delta v \nabla \tilde{Y}_i
    \nabla \tilde{\mu}_i
    \nonumber    
         \\
    &=
  \frac{\tilde{T}}{\hat{\rho}}
    \nabla \cdot \left(
    \alpha_e \Lambda \hat{\rho} \, \delta v 
     \nabla \tilde{s}
    \right)
    +
    \alpha_e \Lambda \, \delta v
    \left(
    \nabla \tilde{s} \nabla \tilde{T}
    +
    \sum_i 
    \nabla \tilde{Y}_i
    \nabla \tilde{\mu}_i
    \right).
    \nonumber    
\end{align}
 Expressed as a Eulerian  equation for $\hat{\rho} \tilde{s}$, this becomes
\begin{equation}
\label{eq:sconv}
\left(\frac{\pd \hat{\rho}\tilde{s}}{\pd t}\right)_\mathrm{conv}=
    \nabla \cdot \left(
    \alpha_e \Lambda \hat{\rho} \, \delta v 
     \nabla \tilde{s}
    \right)
    +
    \frac{\alpha_e \hat{\rho} \Lambda \, \delta v}{\tilde{T}}
    \left(
    \nabla \tilde{s} \nabla \tilde{T}
    +
    \sum_i 
    \nabla \tilde{Y}_i
    \nabla \tilde{\mu}_i
    \right).
\end{equation}
Now consider the entropy change due to the sink term from buoyant driving
(subscript ``buoy'')
in the internal energy equation~(\ref{eq:kuhfuss_energy2}),
\begin{equation}
\left(\frac{\pd \hat{\rho}\tilde{s}}{\pd t}\right)_\mathrm{buoy}
=
\hat{\rho} \left(\frac{\pd \tilde{s}}{\pd t}\right)_\mathrm{buoy}
=
\frac{\hat{\rho}}{\tilde{T}} \left(\frac{\pd \tilde{\epsilon}}{\pd t}\right)_\mathrm{buoy}
=
\frac{\langle \rho' \mathbf{v}''\rangle \cdot \mathbf{g}}{\tilde{T}}
\end{equation}
In order to obtain a form containing quadratic
terms in some gradients of thermodynamic quantities
as in Equation~(\ref{eq:dsdt_conv}), we
assume hydrostatic equilibrium so that
we can replace $\mathbf{g}$ with $\nabla\hat{P}/\hat{\rho}$:
\begin{equation}
\left(\frac{\pd \hat{\rho}\tilde{s}}{\pd t}\right)_\mathrm{buoy}
=
\frac{\langle \rho' \mathbf{v}''\rangle \cdot \nabla \hat{P}} {\hat{\rho} \tilde{T}}
=
\frac{\Lambda\, \delta v } {\hat{\rho} \tilde{T}}
\left[
\left(\frac{\pd \rho}{\pd P}\right)_{s,Y_i}
\nabla \hat{P}-
\nabla \hat{\rho}
\right]
\cdot \nabla \hat{P}
\end{equation}
In fact, there is some motivation for consistently
using $\nabla\hat{P}/\hat{\rho}$ as the effective gravity
for non-hydrostatic flow because
this ensures that the growth rate of turbulent kinetic
energy is compatible with the growth rate
of the compressible Rayleigh-Taylor instability
(for which see, e.g., \citealt{bandiera_84,benz_90,mueller_91,zhou_17a}).

After expanding $\nabla \hat{P}$ in terms of the partial derivatives of
$P$ with respect to $s$, $Y_i$, and $\rho$, the term $\nabla \hat{\rho}$
cancels out, and we obtain
\begin{equation}
\left(\frac{\pd \hat{\rho}\tilde{s}}{\pd t}\right)_\mathrm{buoy}
=
\frac{\Lambda \hat{\rho}\, \delta v }{ \hat{\rho} \tilde{T}}
\left(\frac{\pd \rho}{\pd P}\right)_{s,Y_i}
\left[
\left(\frac{\pd P}{\pd s}\right)_{\rho,Y_i}
\nabla \tilde{s}
+
\sum_i
\left(\frac{\pd P}{\pd Y_i}\right)_{\rho,s}
\nabla \tilde{Y}_i
\right]
\cdot \nabla \hat{P}.
\end{equation}
We can then exploit Maxwell's relations\footnote{
Note that one can use $\pd /\pd \rho^{-1}=-\rho^2 \pd /\pd \rho$
for all derivatives with respect to $\rho^{-1}$.}
$(\pd P/\pd s)_{\rho, Y_i}=
(\pd T/\pd \rho^{-1})_{s, Y_i}$
and
$(\pd P/\pd Y_i)_{\rho, s}=
(\pd \mu_i/\pd \rho^{-1})_{s, Y_i}$
to replace the thermodynamic derivatives of $P$,
\begin{align}
\left(\frac{\pd \hat{\rho}\tilde{s}}{\pd t}\right)_\mathrm{buoy}
&=
-\frac{\Lambda \hat{\rho}\, \delta v} {\tilde{T}}
\left(\frac{\pd \rho}{\pd P}\right)_{s,Y_i}
\bigg[
\nabla \tilde{s}
\left(\frac{\pd T}{\pd \rho} \right)_{s,Y_i}
+
\sum_i
\nabla \tilde{Y}_i
\left(\frac{\pd \mu_i}{\pd \rho} \right)_{s,Y_i}
\bigg]
\cdot  \nabla \hat{P}
\nonumber
\\
&=
-\frac{\Lambda \hat{\rho}\, \delta v} {\tilde{T}}
\bigg[
\nabla \tilde{s}
\left(\frac{\pd T}{\pd P} \right)_{s,Y_i}
+
\sum_i
\nabla \tilde{Y}_i
\left(\frac{\pd \mu_i}{\pd P} \right)_{s,Y_i}
\bigg]
\cdot  \nabla \hat{P}
\end{align}
We can further rewrite this as
\begin{align}
\label{eq:sbuoy}
\left(\frac{\pd \hat{\rho}\tilde{s}}{\pd t}\right)_\mathrm{buoy}
&=
\frac{\Lambda \hat{\rho}\, \delta v} {\tilde{T}}
\Bigg\{
-
\nabla \tilde{s}\cdot
\nabla \tilde{T}
-
\sum_i
\nabla \tilde{Y}_i\cdot
\nabla \tilde{\mu}_i
\\
&
\phantom{=}
+
\nabla \tilde{s} \cdot
\left[
\left(\frac{\pd T}{\pd s}\right)_{P,Y_i}
\nabla \tilde{s} +
\sum_j \left(\frac{\pd T}{\pd Y_i}\right)_{P,s,Y_k,k\ne i}
\nabla \tilde{Y}_j
\right]
\nonumber
\\
&
\phantom{=}
+
\sum_i
\nabla \tilde{Y}_i
\left[
\left(\frac{\pd \mu_i}{\pd s}\right)_{P,Y_i}
\nabla \tilde{s} +
\sum_j
\left(\frac{\pd \mu_i}{\pd Y_j}\right)_{P,s,Y_k,k\ne j}
\nabla \tilde{Y}_j
\right]
\Bigg\}.
\nonumber
\end{align}
 The terms in the
first line cancel the extra source term in 
$\left(\pd \hat{\rho}\tilde{s}/\pd t\right)_\mathrm{conv}$
if and only if $\alpha_e=1$ and $\alpha_X=1$.
Thus, only the terms in the second 
and third line remain as a net source
or sink term for the entropy.
In order to better understand these
terms, we reformulate them in
terms of the MLT entropy
and mole fractions contrasts
$\delta s$ and $\delta Y_e$ by eliminating
$\delta v$ in favor of
the convective turnover time $\tau=\Lambda/\delta v$,
\begin{align}
\label{eq:sbuoy2}
\left(\frac{\pd \hat{\rho}\tilde{s}}{\pd t}\right)_\mathrm{buoy}
&=
\frac{\Lambda \hat{\rho}\, \delta v} {\tilde{T}}
\left(
-
\nabla \tilde{s}\cdot
\nabla \tilde{T}
-
\sum_i
\nabla \tilde{Y}_i\cdot
\nabla \tilde{\mu}_i
\right)
\\
&
\phantom{=}
+\frac{\hat{\rho}\,} {\tau \tilde{T} }
\Bigg[
\left(\frac{\pd T}{\pd s}\right)_{P,Y_i}
\delta s^2 +
\sum_j \left(\frac{\pd T}{\pd Y_i}\right)_{P,s,Y_k,k\ne i}
\delta s\, \delta {Y}_j
\nonumber
\\
&
\phantom{=}
+
\sum_i
\left(\frac{\pd \mu_i}{\pd s}\right)_{P,Y_i}
\delta s \, \delta Y_i+
\sum_{i,j}
\left(\frac{\pd \mu_i}{\pd Y_j}\right)_{P,s,Y_k,k\ne j}
\delta Y_i\, \delta Y_j
\Bigg],
\nonumber
\end{align}
The terms in the second and third line
can be identified as the contribution
of the mixing entropy for material
with $s=\tilde{s}\pm \delta s$
and $Y_i=\tilde{Y}_i \pm \delta Y_i$,
which is liberated per convective
turnover time.  This is demonstrated
in Appendix~\ref{sec:mixing} for a
general multi-species fluid (cp. \S~21 in 
\citealt{landau_5} for the case of
constant composition).

For well-behaved equations of state
the mixing entropy is always positive; thus
the Kuhfuss model actually predicts the correct positive entropy change due to
turbulent mixing. This is particularly
easy to see for the case without composition gradients, in which case
the sum of
Equations~(\ref{eq:sconv}) and (\ref{eq:sbuoy}) reduces to
\begin{equation}
\left(\frac{\pd \hat{\rho}\tilde{s}}{\pd t}\right)_\mathrm{conv}+
\left(\frac{\pd \hat{\rho}\tilde{s}}{\pd t}\right)_\mathrm{buoy}
    =
    \frac{\Lambda \hat{\rho}\, \delta v}{\tilde{T}}
\left(\frac{\pd \tilde{s}}{\pd r}\right)^2
\left(\frac{\pd T}{\pd s}\right)_{P,Y_i}=
  \frac{\Lambda \hat{\rho}\, \delta v}{\tilde{T}}
\left(\frac{\pd \tilde{s}}{\pd r}\right)^2
c_P^{-1}>0,
\end{equation}
where $c_P$ is the heat capacity at constant pressure.

This analysis of the entropy source term
shows that the Kuhfuss model is 
compatible with the second law, 
providing the correct, unavoidable rate of entropy 
generation in the absence of turbulent dissipation 
and turbulent viscosity. Including a
source term for buoyant driving
in the total energy equation therefore also 
overestimates entropy generation by turbulent 
mixing and dissipation. Moreover,
$\alpha_e$ and $\alpha_X$ are actually
not free parameters, they are fixed by
the coefficient of the buoyant driving
term and the requirement of a correct entropy 
generation rate from
turbulent mixing. Only the ``equation
of state'' $P_t(\hat{\rho},\varpi)$ of the 
turbulent ``eddy gas'', $\alpha_\varpi$,
$\alpha_\mu$, and the mixing length parameter
$\alpha$ remain as free parameters (unless
there are further, yet undiscovered consistency requirements).

We need to stress, however, that consistency with energy conservation and with the second law does
not mean that the Kuhfuss model is \emph{accurate}
for the supernova problem. The closure
relations in the model are still based
on an MLT ansatz that may not be valid 
in the regime of strong compressibility
and for the typical flow geometry encountered
in supernova explosions.
Constructing appropriate closures for the turbulent fluctuations that
 include compressibility effects is anything but trivial (see, e.g., 
\citealt{canuto_93}, and  also \citealt{duffell_16} for a thoughtful
 attempt in the case of the compressible Rayleigh-Taylor instability).

\subsection{Numerical Implementation}
\label{sec:numerics}
In order to illustrate the explosion dynamics of 1D models that
incorporate effects of turbulence, we implement
the model from Section~\ref{sec:kuhfuss_favre} in the neutrino radiation
hydrodynamics code \textsc{CoCoNuT-FMT}
\citep{mueller_10,mueller_15a} and run three different variations
of the model for the $20 M_\odot$ progenitor of
\citet{woosley_07}. \textsc{CoCoNuT-FMT} is a finite-volume
code for general relativistic hydrodynamics in
spherical polar coordinates with the xCFC approximation for the 
metric \citep{cordero_09}. It uses higher-order reconstruction
and an approximate Riemann solver combined with 
a stationary, fast multi-group neutrino transport (\textsc{FMT}) scheme
as described in \citet{mueller_15a}.

Since \textsc{CoCoNuT-FMT} is a relativistic code, some care is required
to implement the turbulence model from Section~\ref{sec:minimal_model},
which is formulated in the Newtonian approximation. Since general
relativistic effects are unimportant and velocities are well below
the speed of light in the gain layer as the main region of interest,
one can dispense with a detailed re-derivation of the equations
in full relativity and simply include a few heuristic modifications:
All flux terms pick up an extra factor $N \phi^4$ in terms
of the lapse function $N$ and the conformal factor $\phi$ of
the conformally flat metric, and all conserved quantities pick up
an extra factor $\phi^6$. Moreover, we use the
relativistic expression for the MLT density contrast
$\rho'$ \citep[cp.][]{mueller_13},
\begin{equation}
    \rho' =\Lambda \left(\frac{\pd \hat{\rho} (1+\tilde{\epsilon}/c^2)}{\pd r}-\frac{1}{c_\mathrm{s}^2}\frac{\pd \hat{P}}{\pd r}\right),
\end{equation}
when computing the Brunt-V\"ais\"al\"a frequency and buoyant driving.
Whether one computes the gravitational acceleration simply as the derivative
of the lapse function or includes relativistic correction factors is immaterial
in practice.

The terms in the turbulence model are implemented in an operator-split approach except for the convective
energy flux and the corresponding
turbulent fluxes in the equations
for the mass fractions and the
Favre-averaged
electron fraction $\tilde{Y}_e$; these are
added directly to the hydro fluxes
obtained from the Riemann solver.

The operator-split update of
the turbulent kinetic energy
and the bulk fluid energy and momentum
are staged as follows: We first 
integrate the terms for buoyant energy
generation and dissipation,
\begin{equation}
    \frac{\pd  \varpi}{\pd t}
    =\max(\omega_\mathrm{BV}^2 \Lambda \varpi^{1/2},0)
    -\, \hat{\rho} \frac{\varpi^{3/2}}{\Lambda_\mathrm{diss}}.
\end{equation}
Here we deviate
from 
\citet{kuhfuss_86} and
\citet{wuchterl_98} by introducing
a dissipation length
$\Lambda_\mathrm{diss}$ that
may be different from the mixing length $\Lambda$. The rationale for
this is that the identification 
of  $\Lambda_\mathrm{diss}$ and
$\Lambda$ can lead to excessive
overshoot into the convectively
stable atmosphere of the
proto-neutron star. We limit
the dissipation length by
balancing the turbulent kinetic
energy and the work against
buoyancy for overshooting by
a distance $\Lambda_\mathrm{diss}$,
\begin{equation}
    \Lambda_\mathrm{diss}=
    \max\left[\min\left(\Lambda,
    \sqrt{\frac{\varpi^2}{
    \max(-2\omega_\mathrm{BV}^2,0)}}\right),
    10^{-10}\, \mathrm{cm}\right].
\end{equation}
This is motivated by the realisation that the 
penetration depth of convective plumes from
the gain region is not simply a fixed multiple
of the pressure scale height, but regulated
by the kinetic energy of the overshooting plumes
and the Brunt-V\"ais\"al\"a frequency
in the convectively stable cooling
layer  \citep[][see their Equation 26]{murphy_09}.
This limiting procedure for the dissipation
length is only applied in regions where
$\omega_\mathrm{BV}^2<0$ indicates convective stability.

Since the source term for buoyant
driving vanishes for $\varpi=0$,
care must be taken to ensure that convective
motions can actually grow. One possibility
is to imposes a small seed value for $\varpi$
as done by \citet{couch_19}. We instead circumvent
this problem altogether
by evolving 
$\delta v$, whose evolution equation can
be obtained using the chain rule:
\begin{equation}
    \frac{\pd \delta v}{\pd t}
    =\max\left(\frac{\omega_\mathrm{BV}^2 \Lambda}{2},0\right)
    -\frac{C_\mathrm{D}}{2} \, \hat{\rho} \frac{\varpi^{2}}{\Lambda_\mathrm{diss}}.
\end{equation}
No seed for $\delta v$ is required since $\pd \delta v/\pd t$
is non-zero even for $\delta v=0$. Since this equation is stiff if
$\Lambda_\mathrm{diss}$ is very small,
we solve for the value $\delta v_\mathrm{new}$
at the next time step implicitly,
and afterwards solve
\begin{equation}
\label{eq:energy_update}
    \frac{\pd  \tilde{\epsilon}}{\pd t}
    =-\max(\omega_\mathrm{BV}^2 \Lambda \varpi^{1/2},0)
    +\, \hat{\rho} \frac{\varpi^{3/2}}{\Lambda_\mathrm{diss}}.
\end{equation}
by updating the internal energy as
\begin{eqnarray}
    \tilde{\epsilon}_\mathrm{new}
    =\tilde{\epsilon}_\mathrm{old}
    +({\delta v}_\mathrm{old}^2-{\delta v}_\mathrm{new}^2).
\end{eqnarray}
As discussed in Section~\ref{sec:entropy},
one can argue that $\hat{\rho}^{-1} \pd \hat{P}/\pd r$
should actually replace  $g$
as the effective gravitational acceleration
in the source term for buoyant driving in the
turbulent kinetic energy equation
(\ref{eq:kuhfuss_varpi2}) and in the calculation
of the Brunt~V\"ais\"al\"a frequency. This
would have the virtue that the
direction of unstable gradients
would reverse once the pressure gradient
behind the shock becomes positive as in
the case of Rayleigh-Taylor later during
the explosion 
\citep{chevalier_76,mueller_91,fryxell_91}.
Using $\hat{\rho}^{-1} \pd \hat{P}/\pd r$
as the effective gravity can, however,
create numerical problems at the shock,
and following late-time mixing instabilities
would also require other changes in the
model assumptions, e.g., on the mixing
length \citep{duffell_16}.
For this reason, we still choose
to compute the Brunt-V\"ais\"al\"a frequency
using the gravitational acceleration $g$
instead of $\rho^{-1} \pd\hat{P}/\pd r$.

We next update the fluid velocity according to
\begin{equation}
    \frac{\pd \tilde{v}_r}{\pd t}
    =-\frac{1}{\hat{\rho}}\frac{\pd P_\mathrm{t}}{\pd r},
\end{equation}
and then perform a step for diffusion,
advection, and $P\, \ud V$ work,
\begin{multline}
    \frac{\pd \hat{\rho} \varpi}{\pd t}
    =
    -\frac{1}{r^2}\frac{\pd}{\pd r}
    \left[r^2 \hat{\rho}\tilde{v}_r  \varpi\right]
    -\frac{P_t}{r^2}\frac{\pd r^2 \tilde{v}_r}{\pd r}
    -\frac{\alpha_\varpi}{r^2}\frac{\pd}{\pd r}
    \left(r^2 \hat{\rho} \Lambda \varpi^{1/2} \frac{\pd \varpi}{\pd r}\right).
\end{multline}
Finally, we integrate the turbulent viscosity terms,
\begin{eqnarray}
\frac{\pd \tilde{v}_r}{\pd t}
&=&
\frac{4}{3 \hat{\rho} r^3}\frac{\pd }{\pd r}
\left[r^3 \mu
\left(\frac{\pd \tilde{v}_r}{\pd r}
-\frac{\tilde{v}_r}{r}\right)\right],
\\
\frac{\pd \tilde{\epsilon}}{\pd t}
&=&
\frac{4}{3} \Lambda \varpi^{1/2} \left(\frac{\pd \tilde{v}_r}{\pd r}-\frac{\tilde{v}_r}{r}\right)^2,
\end{eqnarray}
and convert back from the primitive
variables to the conserved variables in
general relativity,

\section{Numerical Results}
\label{sec:results}
We evolve the
$20 M_\odot$ progenitor
of \citet{woosley_07}
until $13\,\mathrm{ms}$
after collapse and then switch on
the turbulence terms. We compare
 three variations of
the turbulence model described
in Sections~\ref{sec:kuhfuss_favre}
and \ref{sec:numerics}:
\begin{itemize}
\item As a baseline model, we consider 
a case where turbulent viscosity
is switched off. 
\item As a second case,
we run another model without
turbulent viscosity, where we 
switch off the sink term 
for buoyant driving in the internal energy equation
at $42 \, \mathrm{ms}$
after bounce.  In other words, instead
of Equation~(\ref{eq:energy_update}), we solve
\begin{equation}
    \frac{\pd  \tilde{\epsilon}}{\pd t}
    = \rho \frac{\varpi^{3/2}}{\Lambda_\mathrm{diss}}
\end{equation}
as in \citet{couch_19}.
This case illustrates
the effect of energy non-conservation.
\item In a third simulation, we revert to
the conservative formulation of buoyant driving and
switch on turbulent viscosity.
\end{itemize}
All the dimensionless
parameters are set to unity in those three
cases, i.e.
$\alpha=\alpha_e=\alpha_X=\alpha_\varpi=\alpha_\mu=C_\mathrm{D}=1$ .

In addition, we also investigate the influence 
of the mixing-length parameter
$\alpha$ in the conservative case
without turbulent viscosity in another
three runs with $\alpha=0.6$,
$\alpha=1.5$, and $\alpha =2$
while keeping $\alpha_\varpi=C_\mathrm{D}=1$
(as well as $\alpha_e=\alpha_X=1$ as dictated
by consistency requirements, see Section~\ref{sec:entropy}).

\begin{figure}
    \centering
    \includegraphics[width=\linewidth]{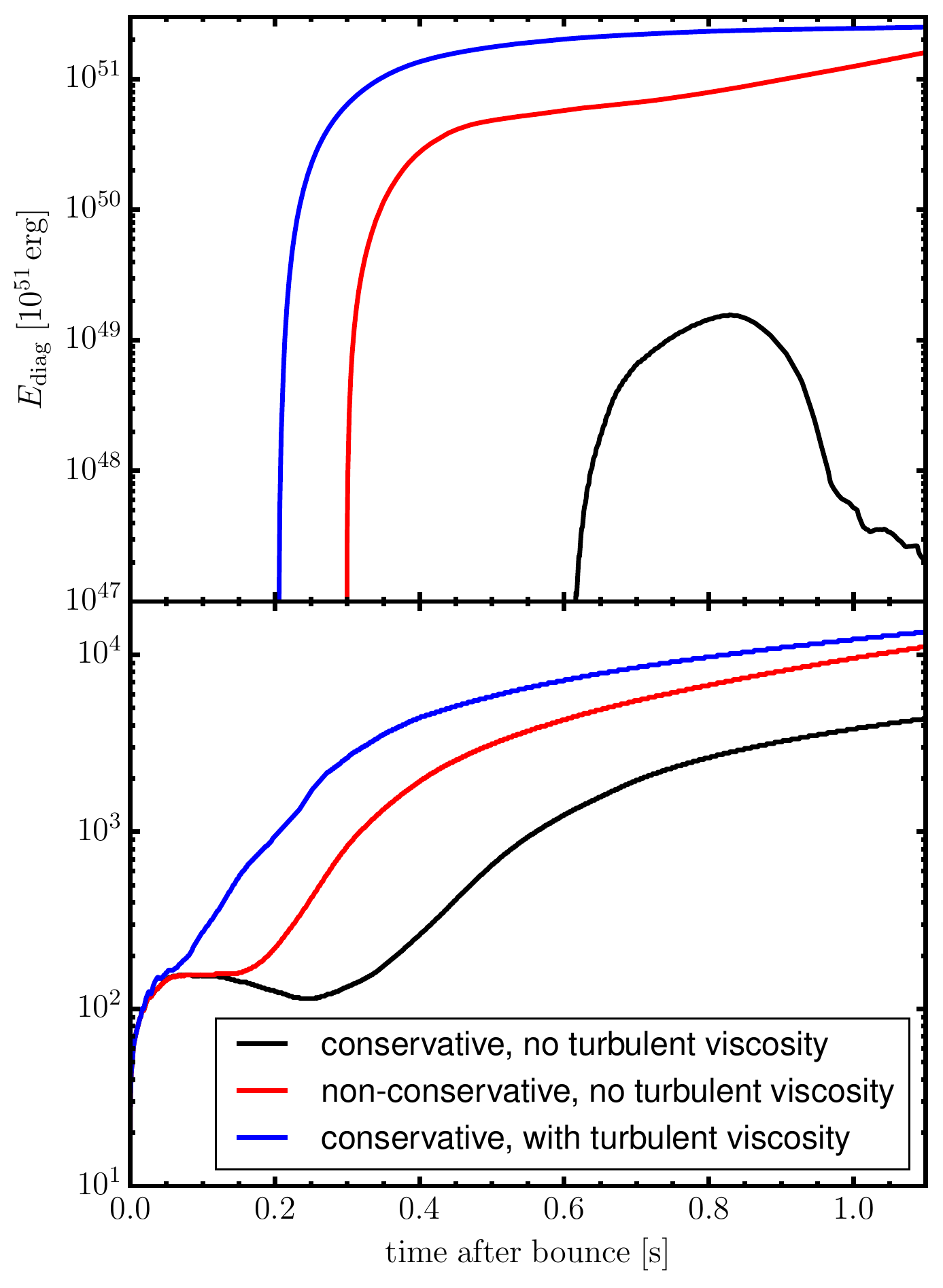}
    \caption{Diagnostic explosion energy
    $E_\mathrm{diag}$ (top panel)
    and shock trajectories (bottom panel) for
    three variations of our 1D turbulence model.
    If the buoyant driving is formulated in
    a conservative manner and turbulent viscosity
    is not included (black curves), shock revival is
    obtained but the explosion fizzles. If energy
    conservation is explicitly violated (red curve), the
    explosion energy reaches $\mathord{\sim}
    2\times 10 ^{51}\, \mathrm{erg}$. The inconsistency
    of this model also shows in an accelerated rate of
    increase of  $E_\mathrm{diag}$ at late times.
    The conservative turbulence model
    with turbulent viscosity produces
    an energetic explosion with an energy of more than
    $2\times 10 ^{51}\, \mathrm{erg}$, but
    the rise of the explosion energy is faster
    than in the sophisticated
    3D model of \citet{melson_15b} of the
    same progenitor, and other considerations
    also suggest that a plausible value
    of the explosion energy is only reached
    by accident (see text).}
    \label{fig:shock_expl}
\end{figure}

Figure~\ref{fig:shock_expl} compares
the shock trajectories and diagnostics explosion energies
$E_\mathrm{diag}$ for the 
first
three cases.
Following the usual definition
\citep{buras_06b,mueller_12a}, we compute $E_\mathrm{diag}$ 
as the total (internal+kinetic+potential) energy of
the material that is nominally unbound at a given time.
The turbulent kinetic energy $\varpi$ is included
in the total energy.

\subsection{Impact of Energy Non-Conservation}
The shock is revived in all three cases, but the explosion
dynamics is significantly different. Comparing
the two runs with and without the sink term for buoyant
driving in the internal energy equation, we find
that ignoring energy conservation in turbulence
models has serious repercussions. Shock revival occurs
more than $100\, \mathrm{ms}$ later for the energy-conserving
turbulence model.

\begin{figure}
    \centering
    \includegraphics[width=\linewidth]{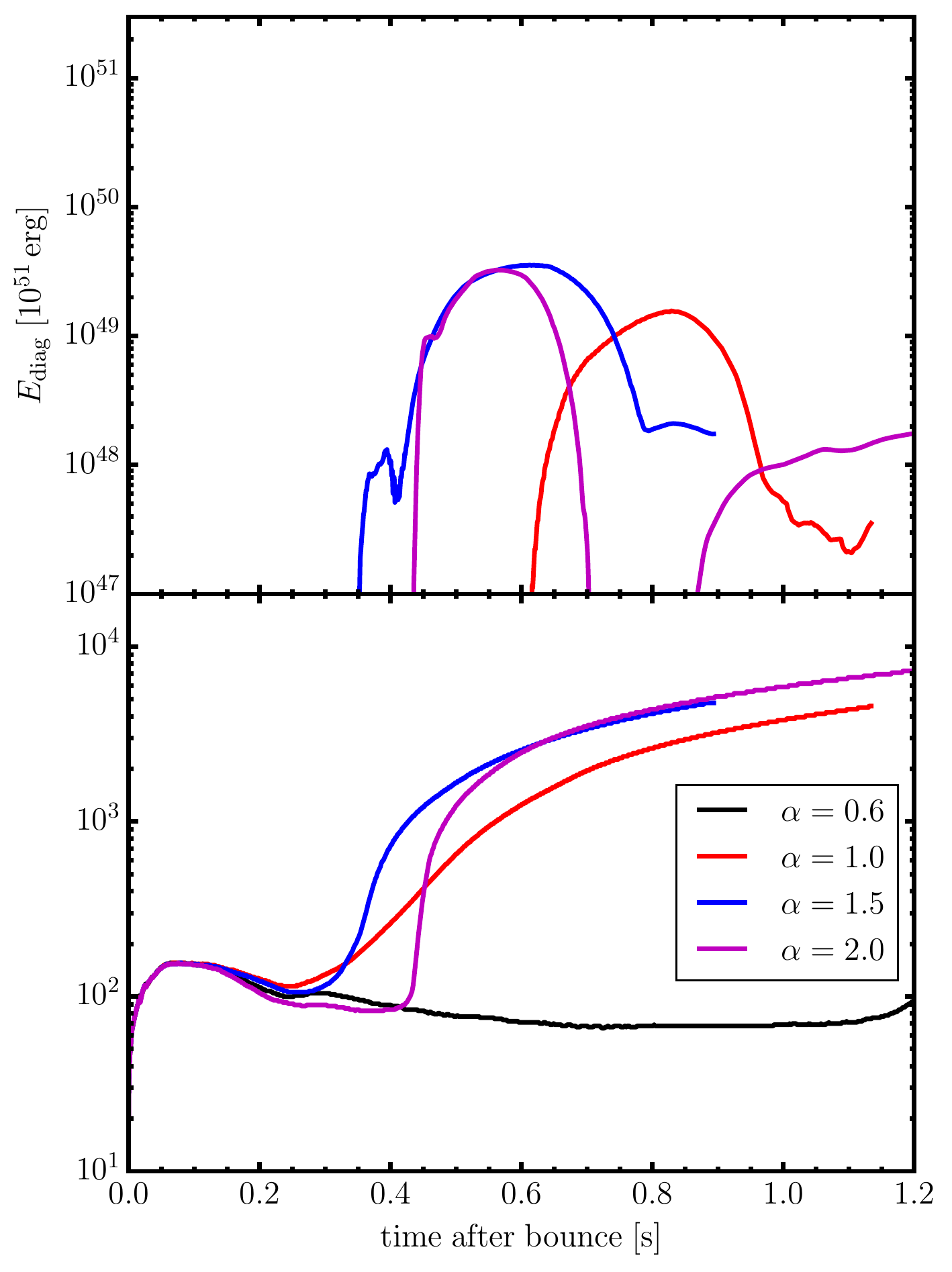}
    \caption{
    Diagnostic explosion energy
    $E_\mathrm{diag}$ (top panel)
    and shock trajectories (bottom panel) for
    the conservative turbulence model
    without turbulent viscosity for
    four different values of the mixing
    length parameter $\alpha$. In all cases,
    the shock is revived, but the explosion
    energy remains very low. Due to feedback
    effects (in particular the decrease
    of the turbulent pressure
    for large $\alpha$), the time of shock revival depends non-montonically on
    $\alpha$.}
    \label{fig:shock_expl_alpha}
\end{figure}

\begin{figure*}
    \centering
    \includegraphics[width=\linewidth]{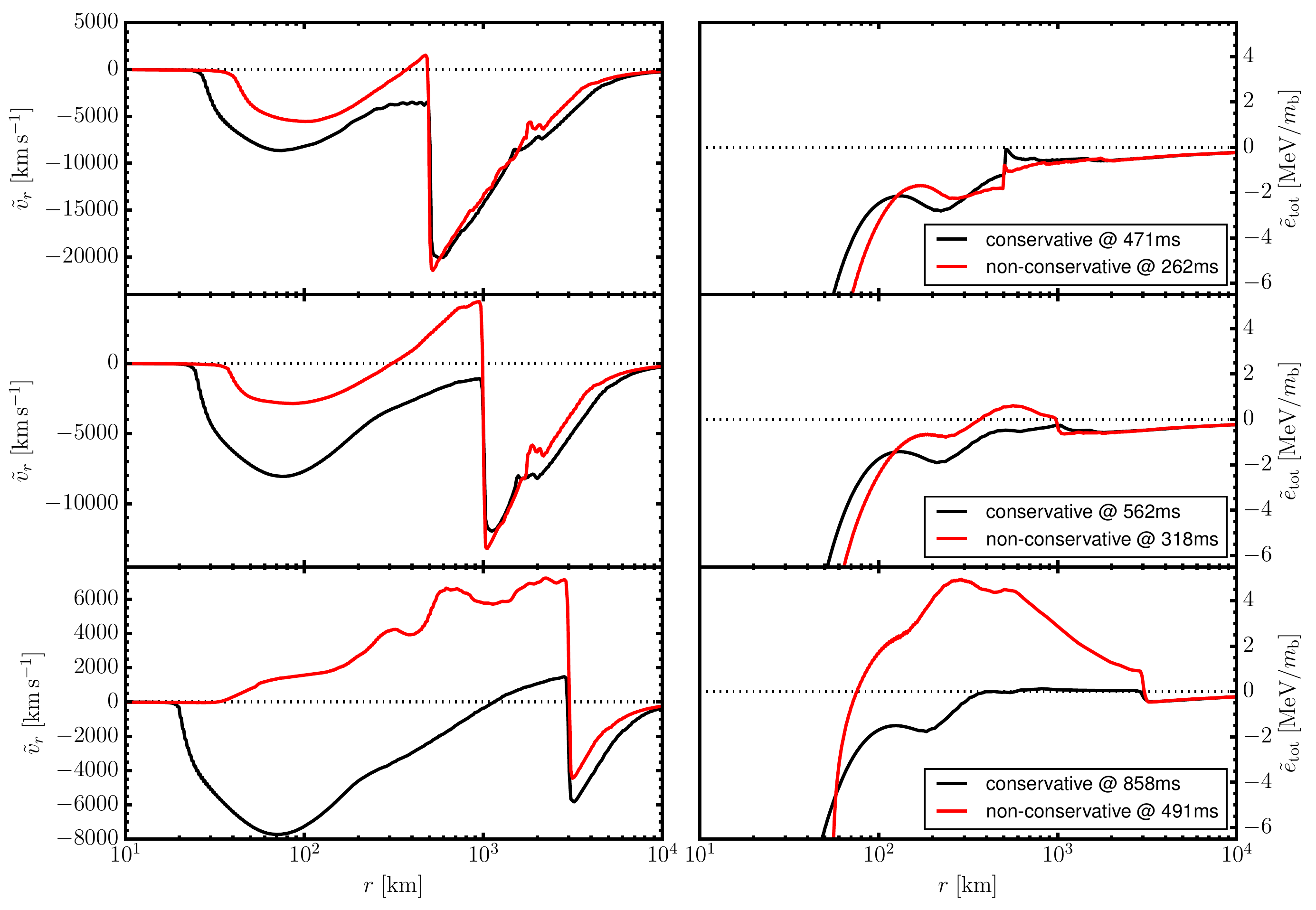}
    \caption{Profiles of the radial velocity $\tilde{v}_r$
    (left column) and the total specific energy
    $\tilde{e}_\mathrm{tot}$
    (excluding rest-mass contributions) in units
    of $\mathrm{MeV}$ per baryon mass $m_\mathrm{b}$
    (right column) for the two runs without turbulent viscosity at three different stages of shock propagation to
    $500 \, \mathrm{km}$,
    $1000 \, \mathrm{km}$,
    and $3000 \, \mathrm{km}$ (top to bottom).
    The post-bounce times at which these
    shock radii are reached are indicated
    in the legends.
    Black and red curves are used for the model
    with a conservative and non-conservative formulation
    of buoyant driving, respectively.
    In the model that respects total energy conservation,
    the post-shock velocity is consistently lower
    and the material in the outer
    part of the post-shock region
    is only marginally unbound.}
    \label{fig:profiles}
\end{figure*}

What is even more concerning, however, is that
the rise of the explosion energy after shock revival
is completely different. With the conservative
formulation of the buoyant driving term,
$E_\mathrm{diag}$ grows slowly
to reach little more than
of $\mathord{\sim}10^{49}\, \mathrm{erg}$.
The diagnostic energy then decreases again so 
that the ejecta are only marginally unbound in the end.
Even though the shock continuously propagates
outward, the result is only a fizzle and not a full-blown explosion.
Figure~\ref{fig:profiles} provides
more explanation for this behaviour by comparing
profiles of the total specific energy
$\tilde{e}_\mathrm{tot}$ and radial velocity
from the conservative and non-conservative model
for similar shock radii.
For a given shock radius, the conservative
model consistently exhibits lower
$\tilde{e}_\mathrm{tot}$ and radial velocity
behind the shock. The turbulent flux terms and the 
turbulent pressure evidently alter the thermodynamic
conditions on the downwind side of the shock
sufficiently to maintain continuous shock expansion,
but without pumping an appreciable amount of energy into the ejecta.

The late-time behaviour of the case with a non-conservative
formulation of the buoyant driving term is
particularly disturbing as it is clearly
unrealistic. Initially, the rise of 
$E_\mathrm{diag}$ slows down 
as physically expected from a declining neutrino 
heating rate with a visible
break around a post-bounce time of $450 \, 
\mathrm{ms}$. A few hundreds of miliseconds later, 
however, $E_\mathrm{diag}$ again starts to 
increase at an \emph{accelerated} rate.

\begin{figure}
    \centering
    \includegraphics[width=\linewidth]{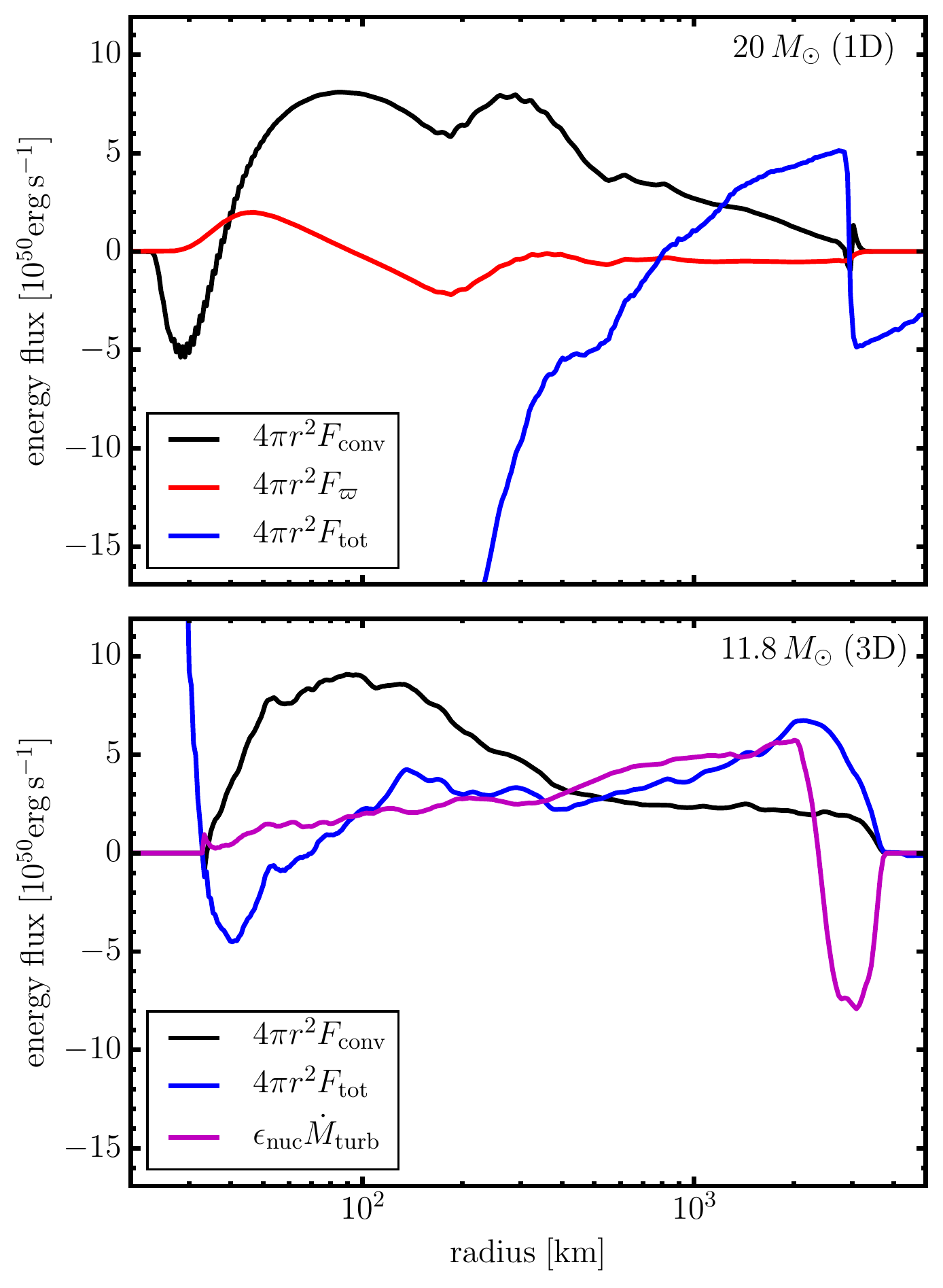}
    \caption{
    Top: convective energy
    flux $F_\mathrm{conv}$ (black), 
    turbulent kinetic
    energy flux
    $F_\varpi$ (red), and 
    total energy flux $F_\mathrm{tot}$ (blue)
    (including turbulent fluctuations
    and the bulk flow)
    in the conservative model without
    turbulent viscosity.
    Bottom:  $F_\mathrm{conv}$ (black)
    and $F_\mathrm{tot}$ (blue) 
    at a post-bounce time for
    a 3D explosion model 
    of an $11.8 M_\odot$
    progenitor \citep{mueller_19}.
    The fluxes for
    the 3D model are shown at a post-bounce time of
    $0.455 \, \mathrm{s}$, where the shock radius
    is about $3000 \, \mathrm{km}$ as in the
    1D model of the $20 M_\odot$ star. At this
    point, the explosion energy of the 3D model
    is about $10^{50}\, \mathrm{erg}$.
    The total energy flux
    is similar to 
    $\epsilon_\mathrm{nuc}\dot{M}_\mathrm{turb}$
    (magenta curve) over a wide range in radii
    assuming a recombination energy
    of $\epsilon_\mathrm{nuc}=5 \, \mathrm{MeV}$.
    }
    \label{fig:fconv}
\end{figure}

These findings lead to two conclusions: First, the question of total
energy conservation clearly must not be ignored when
formulating a 1D turbulence model for supernova explosions;
it is crucial for the energetics of the explosion.
Second, correctly reproducing the energetics within a 1D turbulence model
is evidently a very different matter than merely
reproducing shock revival and shock expansion in a seemingly
realistic way. It is non-trivial to ensure that the shock propagation
is coupled to the explosion energetics in the same manner
as it is in multi-D simulations, where
the shock velocity $v_\mathrm{sh}$ roughly follows
the analytic scaling law derived by \citet{matzner_99},
\begin{equation}
v_\mathrm{sh}
\sim\sqrt{\frac{E_\mathrm{diag}}{M_\mathrm{gain}}}
\left(\frac{M_\mathrm{gain}}{\rho_\mathrm{pre} r^3}\right)^{0.19}
\end{equation}
in terms of $E_\mathrm{diag}$, the mass $M_\mathrm{gain}$ between the shock 
and the gain radius, and the pre-shock density $\rho_\mathrm{pre}$
as shown by \citet{mueller_15b}.
The conservative model without turbulent
viscosity does not ensure this, and is
therefore also unsatisfactory.

One might suspect that the low explosion
energy in the conservative case is
just an accidental pathology due to an unfortunate
choice of the non-dimensional parameters.
Varying the mixing-length parameter
$\alpha$ does not qualitatively change
the behaviour of the model, however.
Figure~\ref{fig:shock_expl_alpha} shows
that the explosion energy remains
unacceptably low even if we vary
$\alpha$ over a wide range from
$\alpha=0.6$ to $\alpha=2$. Whether
different choices for
$\alpha_\varpi$ and $C_\mathrm{D}$
could remedy the problem
of fizzling explosions is a moot point;
if this were the case, one could still
justifiably doubt the robustness of a model
that can easily behave in an unsatisfactory
manner.

As an aside, Figure~\ref{fig:shock_expl_alpha}
reveals that the simulations  do not react monotonically to changes in $\alpha$ if
$\alpha$ is varied over a wider range
than in \citet{couch_19}. The models
with $\alpha=1.5$ and $\alpha=2$ actually
explode later than the one with $\alpha=1$,
but then result in faster shock expansion
once the explosion sets in.
Although this may seem puzzling at first
glance, such a non-monotonic behavior
is not unexpected. As $\alpha$ is increased,
buoyant driving becomes stronger for
the same superadiabatic gradient, but
on the other hand mixing
becomes more effective at flattening
the unstable gradient in the
gain region for a given turbulent
velocity,
which in turn reduces buoyant driving. 
If $\alpha$ is increased sufficiently, the net
result of these two competing effects
can be, at least in certain situations, a decrease of the
turbulent pressure and a smaller shock radius.

\subsection{The Full Model with Turbulent Viscosity and its Limitations}
But could 1D turbulence models fare 
better if they included more turbulent
correlation terms, or if some of
the dimensionless coefficients were adjusted?
The conservative model with 
turbulent viscosity superficially points in this
direction, as it respects
energy conservation and reaches an explosion
energy that is not far from the values
observed for supernovae from
progenitors with massive helium cores like 
SN~1987A 
with $\mathord{\sim}1.5 \times 10^{51}\, \mathrm{erg}$ 
\citep{arnett_89} and Cas~A with $\mathord{\sim}2.3 \times 10^{51}\, \mathrm{erg}$  \citep{orlando_16}.

Nonetheless, this turbulence model also fails
to convince upon closer examination.
The diagnostic energy rises significantly
faster than in the exploding
3D model of \citet{melson_15b} for
the same progenitor, and also faster
than in other 3D explosion models
of progenitors
above $15 M_\odot$ \citep{lentz_15,mueller_17,vartanyan_19}.
One could assume that this is simply the 
consequence of early shock revival in this model,
and might be fixed by a reasonable adjustment
of the dimensionless parameters of the
turbulence model, or by eliminating potential
inconsistencies in the formulation of the
turbulent viscosity term at the shock.

However, if the turbulence model could be
tweaked to achieve more realistic explosion 
dynamics, this would still not imply
that it captures the relevant physics
of multi-D models \emph{consistently}. There are
several outstanding issues that would
still need to be addressed.

In multi-D, the explosion energy
is supplied mostly by nucleon recombination of
neutrino-driven outflows, which contributes
an energy of 
$\epsilon_\mathrm{nuc}=5\texttt{-}8 \, \mathrm{MeV}$ per baryon\footnote{This range of values also accounts for some losses from turbulent energy
transfer between the outflows and downflows \citep{mueller_15b}.}
in the outflows \citep{marek_09,mueller_15b}.
The total enthalpy and mass flux 
in the outflows are remarkably constant
over a wide range in radius; the total 
enthalpy flux in the downflows varies
more strongly, but is a subdominant
contribution to the angle-integrated
total enthalpy flux outside a few hundred
kilometres
(see Figure~18 in \citealp{mueller_15b}). What a turbulence model
would need to reproduce to be viewed
as consistent during the rise phase
of the explosion energy is a convective
total enthalpy flux of $\epsilon_\mathrm{nuc} \dot{M}_\mathrm{out}$
from the recombination radius out to several
thousands of kilometres, where most of the
mass of the accumulated ejecta is located.

Given that turbulence models can produce
convective velocities that match
multi-D simulations reasonably
well (see Figure~1 in \citealt{couch_19}),
one may grant that they can also be
used to roughly predict the
outflow rate as\footnote{
During the rise phase of the explosion energy, one
can neglect 
the spherically averaged velocity  $\tilde{v}_r$ when estimating the
outflow rate because $\tilde{v}_r<c_\mathrm{s}$,
whereas $\delta v =\mathcal{O}(c_\mathrm{s})$.
Once $\tilde{v}_r$ approaches the speed of sound, the
cycle of accretion, neutrino heating, and mass ejection
ceases, and the rise of the explosion energy proceeds
at a slower rate during the subsequent
wind phase, cp.\ \citep{marek_09,mueller_15b}.}
$\dot{M}_\mathrm{out}\sim 2 \pi r^2\hat{\rho}\, \delta v$.
Since the energy carried by the outflows
is split between internal energy
and kinetic energy (whose contribution
is more important at large radii),
there is also some justification for
interpreting the kinetic contribution
to the energy flux into the ejecta
as arising from ``turbulent viscosity''
in the framework of a spherical Reynolds
decomposition, although this interpretation
may not be very intuitive or useful.

But if the plausible energetics
in the run with turbulent viscosity is
more than a lucky coincidence resulting
from a fortunate choice of
non-dimensional model parameters
to ensure that the total
energy flux $F_\mathrm{tot}
=F_\mathrm{conv}+F_\mathrm{visc}+F_\mathrm{\varpi}$ carried by 
the overturn motions is roughly given by
the outflow rate times the recombination energy, e.g.,
\begin{equation}
\label{eq:ftot}
    F_\mathrm{conv}+F_\mathrm{visc}+F_\mathrm{\varpi}
    \sim
        \frac{1}{2}
    \hat{\rho} \,\delta v \epsilon_\mathrm{nuc}
    \sim 
    \frac{\epsilon_\mathrm{nuc} \dot{M}_\mathrm{out}}{4\pi r^2}
\end{equation}
over a wide range of radii outside the
recombination region. This holds quite
well in 3D simulations outside $\approx 200 \, \mathrm{km}$
out to some distance behind the shock
during the explosion phase as exemplified
by Figure~18 in \citet{mueller_15b}.

Since the fluxes
$F_\mathrm{conv}$, $F_\mathrm{visc}$,
and $F_\mathrm{\varpi}$
are computed from \emph{local gradients}
of $\tilde{\epsilon}$, $\hat{P}$, $\tilde{v}_r$, and $\varpi$,  it is hard
to conceive of such a mechanism: Why would 
the gradients adjust themselves in such a 
manner that the combined turbulent energy
flux carries a specific energy per unit mass
that corresponds to the energy liberated
at the recombination radius?

This problem is illustrated further in
Figure~\ref{fig:fconv}, where we
compare $F_\mathrm{conv}$,
$F_\varpi$, and the total
(turbulent+bulk) energy flux
$F_\mathrm{tot}$
of the conservative 1D model
without viscosity to 
 $F_\mathrm{conv}$
 and $F_\mathrm{tot}$ in a
 3D model of a different progenitor
 (the $11.8 M_\odot$ model
 of \citealt{mueller_19}) at a similar
 stage of shock propagation.
 Although the profiles
 of  $F_\mathrm{conv}$ are somewhat
 similar (except for a bump
 around the recombination radius
 and a faster drop towards
 the shock), the total
 energy flux is completely different.
 In the 1D model, $F_\mathrm{tot}$
 is completely dominated by the
 advective bulk energy flux at smaller
 radii. By contrast,
 the total energy flux in the 3D
 model is roughly proportional
 to the turbulent mass flux 
 $\dot{M}_\mathrm{turb}\sim \dot{M}_\mathrm{out}$,
 \begin{equation}
 \label{eq:ftot2}
 4\pi r^2 F_\mathrm{tot}
 \sim \epsilon_\mathrm{nuc}\dot{M}_\mathrm{turb},
 \end{equation}
 which is compatible with
 the reasoning behind Equation~(\ref{eq:ftot}).
 Figure~\ref{fig:fconv} also shows that
 the total energy flux is not dominated
 by any single term in the Favre-averaged
 equations; apparently  several
 turbulent terms and the bulk energy flux
need to add up to achieve
 a total energy flux that
 is roughly set by Equations~(\ref{eq:ftot})
 and (\ref{eq:ftot2}). Since a more detailed
 analysis of the 3D explosion models
 in the framework of Favre
 decomposition is beyond the scope of the
 current paper, it remains obscure whether
 and how this can be mirrored in a
 1D turbulence model.

This particular issue is, of course,
only a part of a larger problem that needs
to be investigated further: Although
\citet{murphy_11,murphy_12,mabanta_18}
have studied the applicability of closure
relations for the turbulent correlation
terms in the Reynolds-averaged hydro
equations during the
pre-explosion phase, closures valid
for subsonic convection on the background
of the quasi-stationary accretion flow
before shock revival may no longer be
applicable during the explosion phase.
With a non-stationary background flow,
large-scale overturn motions over
many pressure scale heights, and the 
emergence of supersonic downflows
(which clearly renders the anelastic
approximation invalid), one should
\emph{expect} that the closure relations
and non-dimensional parameters in the 
MLT fluxes  need to change significantly 
during the explosion phase.
For example, one likely needs to  choose
$\alpha \gg 1$ in the definition of
the mixing length $\Lambda=\alpha P/(\rho g)$
during the explosion phase as the radial correlation
length of the convective flow structures increases.
To make matters worse, even the direction of unstable gradients
changes already during the first
seconds as a positive pressure
gradient develops behind the shock,
so that $\emph{low-entropy}$ material
becomes susceptible to being mixed
outward by the Rayleigh-Taylor instability
\citep{chevalier_76,mueller_91,fryxell_91}.
Rather than providing a consistent recipe
for 1D supernova simulations, the seeming ``success''
of the run with turbulent viscosity in fact underscores
all of these concerns: It demonstrates
that \emph{including additional turbulent correlation
terms in the turbulence model can have a significant impact}
because the usual importance hierarchy of the correlation terms
for low-Mach number convection no longer holds around shock revival
and during the explosion.

Another lingering consistency
problem concerns the explosive 
nucleosynthesis during the early
explosion phase. In multi-D
models, a sizeable fraction of the
material synthesised by explosive
burning in the shock is not entrained
by the expanding neutrino-heated bubbles,
but channelled
into accretion downflows and
not ejected \citep{mueller_17,harris_17}. Capturing
this process is crucial for consistently 
predicting the mass of ${}^{56} 
\mathrm{Ni}$ made in the explosion, but
may be inherently beyond 1D turbulence models.
Again, the problem lies in the computation
of turbulent fluxes from local gradients.
In reality, the outflows
should carry a mass flux of ${}^{56}\mathrm{Ni}$
given by
$X_\mathrm{Ni} \dot{M}_\mathrm{out}$, where
$X_\mathrm{Ni}$ is the freeze-out mass fraction
of ${}^{56}\mathrm{Ni}$, which depends
on the entropy
and expansion time-scale of the outflows
\emph{around the freeze-out temperature from nuclear
statistical equilibrium (NSE)}. It is doubtful
that this can be captured accurately by 1D turbulence
models with high accuracy. They might predict
a significant mass flux of  ${}^{56}\mathrm{Ni}$
around the recombination radius, where the gradients
of the mass fractions are steep, but not at larger radii.
Moreover, 1D turbulence models implicitly assume instantaneous
mixing between high-entropy outflows and low-entropy downflows,
which makes the NSE abundances and the results of explosive
burning in the ejected material quite dubious.

\section{Conclusions}
\label{sec:conclusions}
Prompted by the recent works of \citet{mabanta_19}
and \citet{couch_19}, we analysed the consistency of
various 1D turbulence models for 1D supernova simulations
and further bolstered this analysis by numerical experiments.
Our analysis shows that the turbulence models of 
\citet{mabanta_18} and \citet{couch_19} implement
buoyant driving of convection in a manner that
violates energy conservation.
This is because they do not treat the turbulent mass
flux consistently, omitting it in the continuity
equation while including it in the source term for
buoyant energy generation. It is important to stress
that this problem is not one of correct numerical
discretisation; the non-conservation of energy is built
into the analytic form of the model equations.
We also point out that
considerable care must be exercised when formulating
the turbulent flux terms in order to ensure
consistency between the energy and momentum equation
(and the turbulent kinetic energy equation
in time-dependent turbulence models), 
and the correct direction of the convective energy flux.

We point out that the energy-conserving turbulence model of 
\citet{kuhfuss_86} and \citet{wuchterl_98} already fixes the 
inconsistencies in the work of \citet{mabanta_19}
and \citet{couch_19}. Energy conservation is
accomplished by including a sink term for
buoyant driving in the internal energy equation in
this model.
Although this sink may appear unintuitive at first glance,
it emerges consistently from a spherical Favre decomposition,
and ensures the correct evolution of the entropy 
in agreement with the second law of thermodynamics together
with the convective flux terms.
We
have implemented the model of \citet{kuhfuss_86} in the
\textsc{CoCoNuT-FMT} neutrino hydrodynamics code with some minor 
modifications to avoid excessive convective overshoot
and spurious generation of turbulent kinetic energy
in non-homologously expanding or contracting flows.

We simulated the collapse and explosion of
a $20 M_\odot$ progenitor using three different
variations of this 1D turbulence model
in \textsc{CoCoNuT-FMT} to further illustrate
the pitfalls and limitations of this approach.
We find that including buoyant driving as an energy
source term in a non-conservative manner without a 
corresponding sink term dramatically alters the explosion
dynamics. Using the conservative model, the material
behind the continuously expanding shock is barely unbound with
a final explosion energy of less than $10^{48}\, \mathrm{erg}$
as opposed to $\mathord{\sim} 2 \times 10^{51}\, \mathrm{erg}$ with the non-conservative turbulence model.
This result suggests that non-conservative models 
as proposed by \citet{mabanta_19} and \citet{couch_19} should be used with considerable caution. Whether the non-conservative formulation
of buoyant driving not only affects the dynamics of the explosion
phase but also the systematics of ``explodability'', 
which appears rather different in \citet{couch_19} than in
the parameterised models of \citet{ugliano_12,ertl_15,sukhbold_16,mueller_16a,ebinger_19},
also needs to be examined further in the future.

The low explosion energy obtained with our energy-conserving
baseline model demonstrates another consistency
problem of 1D turbulence models. Even if a turbulence model 
accurately captures the point of shock revival 
and predicts plausible shock trajectories, the energetics
of the explosion may still be woefully off. A variation
of the turbulence model including turbulent viscosity
results in a plausible explosion energy for
the $20 M_\odot$ progenitor, but we view this as little
more than a coincidence at this stage. It 
does not imply that 1D turbulence models can 
consistently capture the essential physics that governs the
energetics of supernova explosions in multi-D. 

Before 1D turbulence models can be considered
significantly more consistent
that other phenomenological approaches to
the supernova progenitor-explosion connection
\citep{ugliano_12,pejcha_15a,perego_15,sukhbold_16,mueller_16a},
many critical issues still need to be addressed. Most importantly, one needs to account for
\begin{enumerate}
    \item the coupling of the explosion energy
    to the recombination energy of the neutrino-heated ejecta,
    \item the violation of the anelastic approximation due to the emergence of high-Mach number flow,
    \item changes in the radial correlation length
    of turbulent fluctuations during the developing explosion,
    \item the violation of the hydrostatic approximation and the emergence of an inverse pressure gradient
    behind the shock,
    \item the incomplete entrainment of $^{56}\mathrm{Ni}$ from explosive burning into the neutrino-heated ejecta.
\end{enumerate}
None of the available turbulence models adequately deals with 
these daunting challenges yet. If there is a solution -- which
is not to be taken
for granted -- it will likely not materialise
in the near future and will require a considerably more thorough
analysis of multi-D explosion models than has been carried out 
so far. Because of the complexity of the problem, we cannot
even hope to outline such a solution at this point.

On the other hand, the consistency problems and pitfalls 
that we pointed out should not lead to undue pessimism either.
Even though  1D turbulence models have a long way to go before
they can become a superior, more consistent method
to predict supernova explosion and compact remnant properties
than other approaches, they may complement other
phenomenological supernova models and prove particularly useful
for specific aspects of the progenitor-explosion connection as 
many of the other methods have. 
It is also noteworthy
that the problems pointed out in this work are most acute
during the explosion phase, so that 1D turbulence models
may at least provide a more consistent approach
for studying the conditions for shock revival in 1D across
a wide range of stellar parameters.
In order for
1D turbulence models to find their proper place, it is 
necessary, however, to investigate and incorporate the 
consistency requirements that follow from general physical
principles and supernova explosion physics, to which the 
present work will hopefully contribute.

\section*{Acknowledgements}
%\acknowledgements
I acknowledge fruitful discussions with H.-Th.~Janka, T.~Ertl,
A.~Heger, Q.~Mabanta, and J.~Murphy.  This work was supported by the
Australian Research Council through ARC Future Fellowship
FT160100035. This research was undertaken with the assistance of
resources from the National Computational Infrastructure (NCI), which
is supported by the Australian Government and was supported by
resources provided by the Pawsey Supercomputing Centre with funding
from the Australian Government and the Government of Western
Australia.

\bibliography{paper}

\begin{thebibliography}{}
\makeatletter
\relax
\def\mn@urlcharsother{\let\do\@makeother \do\$\do\&\do\#\do\^\do\_\do\%\do\~}
\def\mn@doi{\begingroup\mn@urlcharsother \@ifnextchar [ {\mn@doi@}
  {\mn@doi@[]}}
\def\mn@doi@[#1]#2{\def\@tempa{#1}\ifx\@tempa\@empty \href
  {http://dx.doi.org/#2} {doi:#2}\else \href {http://dx.doi.org/#2} {#1}\fi
  \endgroup}
\def\mn@eprint#1#2{\mn@eprint@#1:#2::\@nil}
\def\mn@eprint@arXiv#1{\href {http://arxiv.org/abs/#1} {{\tt arXiv:#1}}}
\def\mn@eprint@dblp#1{\href {http://dblp.uni-trier.de/rec/bibtex/#1.xml}
  {dblp:#1}}
\def\mn@eprint@#1:#2:#3:#4\@nil{\def\@tempa {#1}\def\@tempb {#2}\def\@tempc
  {#3}\ifx \@tempc \@empty \let \@tempc \@tempb \let \@tempb \@tempa \fi \ifx
  \@tempb \@empty \def\@tempb {arXiv}\fi \@ifundefined
  {mn@eprint@\@tempb}{\@tempb:\@tempc}{\expandafter \expandafter \csname
  mn@eprint@\@tempb\endcsname \expandafter{\@tempc}}}

\bibitem[\protect\citeauthoryear{{Arnett}, {Bahcall}, {Kirshner}  \&
  {Woosley}}{{Arnett} et~al.}{1989}]{arnett_89}
{Arnett} W.~D.,  {Bahcall} J.~N.,  {Kirshner} R.~P.,   {Woosley} S.~E.,  1989,
  \mn@doi [\araa] {10.1146/annurev.aa.27.090189.003213}, \href
  {http://adsabs.harvard.edu/abs/1989ARA%26A..27..629A} {27, 629}

\bibitem[\protect\citeauthoryear{{Bandiera}}{{Bandiera}}{1984}]{bandiera_84}
{Bandiera} R.,  1984, \aap, \href
  {http://adsabs.harvard.edu/abs/1984A%26A...139..368B} {139, 368}

\bibitem[\protect\citeauthoryear{{Benz} \& {Thielemann}}{{Benz} \&
  {Thielemann}}{1990}]{benz_90}
{Benz} W.,  {Thielemann} F.-K.,  1990, \mn@doi [\apjl] {10.1086/185620}, \href
  {http://adsabs.harvard.edu/abs/1990ApJ...348L..17B} {348, L17}

\bibitem[\protect\citeauthoryear{{Biermann}}{{Biermann}}{1932}]{biermann_32}
{Biermann} L.,  1932, \zap, \href
  {http://adsabs.harvard.edu/abs/1932ZA......5..117B} {5, 117}

\bibitem[\protect\citeauthoryear{{Blondin}, {Mezzacappa}  \&
  {DeMarino}}{{Blondin} et~al.}{2003}]{blondin_03}
{Blondin} J.~M.,  {Mezzacappa} A.,   {DeMarino} C.,  2003, \mn@doi [\apj]
  {10.1086/345812}, \href {http://adsabs.harvard.edu/abs/2003ApJ...584..971B}
  {584, 971}

\bibitem[\protect\citeauthoryear{{B{\"o}hm-Vitense}}{{B{\"o}hm-Vitense}}{1958}]{boehm_58}
{B{\"o}hm-Vitense} E.,  1958, \zap, \href
  {http://adsabs.harvard.edu/abs/1958ZA.....46..108B} {46, 108}

\bibitem[\protect\citeauthoryear{{Buras}, {Janka}, {Rampp}  \&
  {Kifonidis}}{{Buras} et~al.}{2006}]{buras_06b}
{Buras} R.,  {Janka} H.-T.,  {Rampp} M.,   {Kifonidis} K.,  2006, \mn@doi
  [\aap] {10.1051/0004-6361:20054654}, \href
  {http://adsabs.harvard.edu/abs/2006A26A...457..281B} {457, 281}

\bibitem[\protect\citeauthoryear{{Burrows} \& {Goshy}}{{Burrows} \&
  {Goshy}}{1993}]{burrows_93}
{Burrows} A.,  {Goshy} J.,  1993, \mn@doi [\apjl] {10.1086/187074}, \href
  {http://adsabs.harvard.edu/abs/1993ApJ...416L..75B} {416, L75+}

\bibitem[\protect\citeauthoryear{{Burrows}, {Hayes}  \& {Fryxell}}{{Burrows}
  et~al.}{1995}]{burrows_95}
{Burrows} A.,  {Hayes} J.,   {Fryxell} B.~A.,  1995, \mn@doi [\apj]
  {10.1086/176188}, \href {http://adsabs.harvard.edu/abs/1995ApJ...450..830B}
  {450, 830}

\bibitem[\protect\citeauthoryear{{Canuto}}{{Canuto}}{1993}]{canuto_93}
{Canuto} V.~M.,  1993, \mn@doi [\apj] {10.1086/173238}, \href
  {http://adsabs.harvard.edu/abs/1993ApJ...416..331C} {416, 331}

\bibitem[\protect\citeauthoryear{{Canuto} \& {Dubovikov}}{{Canuto} \&
  {Dubovikov}}{1998}]{canuto_98}
{Canuto} V.~M.,  {Dubovikov} M.,  1998, \mn@doi [\apj] {10.1086/305141}, \href
  {http://adsabs.harvard.edu/abs/1998ApJ...493..834C} {493, 834}

\bibitem[\protect\citeauthoryear{{Chevalier}}{{Chevalier}}{1976}]{chevalier_76}
{Chevalier} R.~A.,  1976, \mn@doi [\apj] {10.1086/154557}, \href
  {http://adsabs.harvard.edu/abs/1976ApJ...207..872C} {207, 872}

\bibitem[\protect\citeauthoryear{{Cordero-Carri{\'o}n}, {Cerd{\'a}-Dur{\'a}n},
  {Dimmelmeier}, {Jaramillo}, {Novak}  \& {Gourgoulhon}}{{Cordero-Carri{\'o}n}
  et~al.}{2009}]{cordero_09}
{Cordero-Carri{\'o}n} I.,  {Cerd{\'a}-Dur{\'a}n} P.,  {Dimmelmeier} H.,
  {Jaramillo} J.~L.,  {Novak} J.,   {Gourgoulhon} E.,  2009, \mn@doi [\prd]
  {10.1103/PhysRevD.79.024017}, \href
  {http://adsabs.harvard.edu/abs/2009PhRvD..79b4017C} {79, 024017}

\bibitem[\protect\citeauthoryear{{Couch}, {Warren}  \& {O'Connor}}{{Couch}
  et~al.}{2019}]{couch_19}
{Couch} S.~M.,  {Warren} M.~L.,   {O'Connor} E.~P.,  2019, preprint, \href
  {http://adsabs.harvard.edu/abs/2019arXiv190201340C} {} (\mn@eprint {arXiv}
  {1902.01340})

\bibitem[\protect\citeauthoryear{{Duffell}}{{Duffell}}{2016}]{duffell_16}
{Duffell} P.~C.,  2016, \mn@doi [\apj] {10.3847/0004-637X/821/2/76}, \href
  {http://adsabs.harvard.edu/abs/2016ApJ...821...76D} {821, 76}

\bibitem[\protect\citeauthoryear{{Ebinger}, {Curtis}, {Fr{\"o}hlich}, {Hempel},
  {Perego}, {Liebend{\"o}rfer}  \& {Thielemann}}{{Ebinger}
  et~al.}{2019}]{ebinger_19}
{Ebinger} K.,  {Curtis} S.,  {Fr{\"o}hlich} C.,  {Hempel} M.,  {Perego} A.,
  {Liebend{\"o}rfer} M.,   {Thielemann} F.-K.,  2019, \mn@doi [\apj]
  {10.3847/1538-4357/aae7c9}, \href
  {http://adsabs.harvard.edu/abs/2019ApJ...870....1E} {870, 1}

\bibitem[\protect\citeauthoryear{{Eggleton}}{{Eggleton}}{1983}]{eggleton_83}
{Eggleton} P.~P.,  1983, \mn@doi [\mnras] {10.1093/mnras/204.2.449}, \href
  {http://adsabs.harvard.edu/abs/1983MNRAS.204..449E} {204, 449}

\bibitem[\protect\citeauthoryear{{Ertl}, {Janka}, {Woosley}, {Sukhbold}  \&
  {Ugliano}}{{Ertl} et~al.}{2016}]{ertl_15}
{Ertl} T.,  {Janka} H.-T.,  {Woosley} S.~E.,  {Sukhbold} T.,   {Ugliano} M.,
  2016, \mn@doi [\apj] {10.3847/0004-637X/818/2/124}, \href
  {http://adsabs.harvard.edu/abs/2016ApJ...818..124E} {818, 124}

\bibitem[\protect\citeauthoryear{{Favre}}{{Favre}}{1965}]{favre_65}
{Favre} A.~J.,  1965, \mn@doi [Journal of Applied Mechanics]
  {10.1115/1.3625792}, \href
  {http://adsabs.harvard.edu/abs/1965JAM....32..241F} {32, 241}

\bibitem[\protect\citeauthoryear{{Foglizzo}, {Galletti}, {Scheck}  \&
  {Janka}}{{Foglizzo} et~al.}{2007}]{foglizzo_07}
{Foglizzo} T.,  {Galletti} P.,  {Scheck} L.,   {Janka} H.-T.,  2007, \mn@doi
  [\apj] {10.1086/509612}, \href
  {http://adsabs.harvard.edu/abs/2007ApJ...654.1006F} {654, 1006}

\bibitem[\protect\citeauthoryear{{Foglizzo} et~al.,}{{Foglizzo}
  et~al.}{2015}]{foglizzo_15}
{Foglizzo} T.,  et~al., 2015, \mn@doi [\pasa] {10.1017/pasa.2015.9}, \href
  {http://adsabs.harvard.edu/abs/2015PASA...32....9F} {32, e009}

\bibitem[\protect\citeauthoryear{{Fryxell}, {Arnett}  \& {Mueller}}{{Fryxell}
  et~al.}{1991}]{fryxell_91}
{Fryxell} B.,  {Arnett} D.,   {Mueller} E.,  1991, \mn@doi [\apj]
  {10.1086/169657}, \href {http://adsabs.harvard.edu/abs/1991ApJ...367..619F}
  {367, 619}

\bibitem[\protect\citeauthoryear{{Gehmeyr} \& {Winkler}}{{Gehmeyr} \&
  {Winkler}}{1992}]{gehmeyr_91}
{Gehmeyr} M.,  {Winkler} K.-H.~A.,  1992, \aap, \href
  {http://adsabs.harvard.edu/abs/1992A%26A...253...92G} {253, 92}

\bibitem[\protect\citeauthoryear{{Harris}, {Hix}, {Chertkow}, {Lee}, {Lentz}
  \& {Messer}}{{Harris} et~al.}{2017}]{harris_17}
{Harris} J.~A.,  {Hix} W.~R.,  {Chertkow} M.~A.,  {Lee} C.~T.,  {Lentz} E.~J.,
   {Messer} O.~E.~B.,  2017, \mn@doi [\apj] {10.3847/1538-4357/aa76de}, \href
  {http://adsabs.harvard.edu/abs/2017ApJ...843....2H} {843, 2}

\bibitem[\protect\citeauthoryear{{Herant}, {Benz}, {Hix}, {Fryer}  \&
  {Colgate}}{{Herant} et~al.}{1994}]{herant_94}
{Herant} M.,  {Benz} W.,  {Hix} W.~R.,  {Fryer} C.~L.,   {Colgate} S.~A.,
  1994, \mn@doi [\apj] {10.1086/174817}, \href
  {http://adsabs.harvard.edu/abs/1994ApJ...435..339H} {435, 339}

\bibitem[\protect\citeauthoryear{{H\"udepohl}}{{H\"udepohl}}{2014}]{huedepohl_phd}
{H\"udepohl} L.,  2014, PhD thesis, Technische Universit{\"a}t M{\"u}nchen

\bibitem[\protect\citeauthoryear{{Janka}}{{Janka}}{2012}]{janka_12}
{Janka} H.-T.,  2012, \mn@doi [Annual Review of Nuclear and Particle Science]
  {10.1146/annurev-nucl-102711-094901}, \href
  {http://adsabs.harvard.edu/abs/2012ARNPS..62..407J} {62, 407}

\bibitem[\protect\citeauthoryear{{Janka} \& {M\"uller}}{{Janka} \&
  {M\"uller}}{1995}]{janka_95}
{Janka} H.-T.,  {M\"uller} E.,  1995, \mn@doi [\apjl] {10.1086/309604}, \href
  {http://adsabs.harvard.edu/abs/1995ApJ...448L.109J} {448, L109}

\bibitem[\protect\citeauthoryear{{Kitaura}, {Janka}  \&
  {Hillebrandt}}{{Kitaura} et~al.}{2006}]{kitaura_06}
{Kitaura} F.~S.,  {Janka} H.-T.,   {Hillebrandt} W.,  2006, \mn@doi [\aap]
  {10.1051/0004-6361:20054703}, \href
  {http://adsabs.harvard.edu/abs/2006A26A...450..345K} {450, 345}

\bibitem[\protect\citeauthoryear{{Kuhfuss}}{{Kuhfuss}}{1986}]{kuhfuss_86}
{Kuhfuss} R.,  1986, \aap, \href
  {http://adsabs.harvard.edu/abs/1986A%26A...160..116K} {160, 116}

\bibitem[\protect\citeauthoryear{{Landau} \& {Lifschitz}}{{Landau} \&
  {Lifschitz}}{1969}]{landau_5}
{Landau} L.~D.,  {Lifschitz} E.~M.,  1969, {Course of theoretical physics}.
 Vol. V, Statistical Physics, Part I, Pergamon Press, Oxford

\bibitem[\protect\citeauthoryear{{Lentz} et~al.,}{{Lentz}
  et~al.}{2015}]{lentz_15}
{Lentz} E.~J.,  et~al., 2015, \mn@doi [\apjl] {10.1088/2041-8205/807/2/L31},
  \href {http://adsabs.harvard.edu/abs/2015ApJ...807L..31L} {807, L31}

\bibitem[\protect\citeauthoryear{{Mabanta} \& {Murphy}}{{Mabanta} \&
  {Murphy}}{2018}]{mabanta_18}
{Mabanta} Q.~A.,  {Murphy} J.~W.,  2018, \mn@doi [\apj]
  {10.3847/1538-4357/aaaec7}, \href
  {http://adsabs.harvard.edu/abs/2018ApJ...856...22M} {856, 22}

\bibitem[\protect\citeauthoryear{{Mabanta}, {Murphy}  \& {Dolence}}{{Mabanta}
  et~al.}{2019}]{mabanta_19}
{Mabanta} Q.~A.,  {Murphy} J.~W.,   {Dolence} J.~C.,  2019, preprint, \href
  {http://adsabs.harvard.edu/abs/2019arXiv190111234M} {} (\mn@eprint {arXiv}
  {1901.11234})

\bibitem[\protect\citeauthoryear{{Marek} \& {Janka}}{{Marek} \&
  {Janka}}{2009}]{marek_09}
{Marek} A.,  {Janka} H.,  2009, \mn@doi [\apj] {10.1088/0004-637X/694/1/664},
  \href {http://adsabs.harvard.edu/abs/2009ApJ...694..664M} {694, 664}

\bibitem[\protect\citeauthoryear{{Matzner} \& {McKee}}{{Matzner} \&
  {McKee}}{1999}]{matzner_99}
{Matzner} C.~D.,  {McKee} C.~F.,  1999, \mn@doi [\apj] {10.1086/306571}, \href
  {http://adsabs.harvard.edu/abs/1999ApJ...510..379M} {510, 379}

\bibitem[\protect\citeauthoryear{{Mayle}}{{Mayle}}{1985}]{mayle_85}
{Mayle} R.~W.,  1985, PhD thesis, California Univ., Berkeley.

\bibitem[\protect\citeauthoryear{{Melson}, {Janka}, {Bollig}, {Hanke}, {Marek}
  \& {M{\"u}ller}}{{Melson} et~al.}{2015}]{melson_15b}
{Melson} T.,  {Janka} H.-T.,  {Bollig} R.,  {Hanke} F.,  {Marek} A.,
  {M{\"u}ller} B.,  2015, \mn@doi [\apjl] {10.1088/2041-8205/808/2/L42}, \href
  {http://adsabs.harvard.edu/abs/2015ApJ...808L..42M} {808, L42}

\bibitem[\protect\citeauthoryear{{Mirizzi}, {Tamborra}, {Janka}, {Saviano},
  {Scholberg}, {Bollig}, {H{\"u}depohl}  \& {Chakraborty}}{{Mirizzi}
  et~al.}{2016}]{mirizzi_16}
{Mirizzi} A.,  {Tamborra} I.,  {Janka} H.-T.,  {Saviano} N.,  {Scholberg} K.,
  {Bollig} R.,  {H{\"u}depohl} L.,   {Chakraborty} S.,  2016, \mn@doi [Nuovo
  Cimento Rivista Serie] {10.1393/ncr/i2016-10120-8}, \href
  {http://adsabs.harvard.edu/abs/2016NCimR..39....1M} {39, 1}

\bibitem[\protect\citeauthoryear{{M{\"u}ller}}{{M{\"u}ller}}{2015}]{mueller_15b}
{M{\"u}ller} B.,  2015, \mn@doi [\mnras] {10.1093/mnras/stv1611}, \href
  {http://adsabs.harvard.edu/abs/2015MNRAS.453..287M} {453, 287}

\bibitem[\protect\citeauthoryear{{M{\"u}ller} \& {Janka}}{{M{\"u}ller} \&
  {Janka}}{2015}]{mueller_15a}
{M{\"u}ller} B.,  {Janka} H.-T.,  2015, \mn@doi [\mnras]
  {10.1093/mnras/stv101}, \href
  {http://adsabs.harvard.edu/abs/2015MNRAS.448.2141M} {448, 2141}

\bibitem[\protect\citeauthoryear{{M\"uller}, {Fryxell}  \& {Arnett}}{{M\"uller}
  et~al.}{1991}]{mueller_91}
{M\"uller} E.,  {Fryxell} B.,   {Arnett} D.,  1991, \aap, \href
  {http://adsabs.harvard.edu/abs/1991A%26A...251..505M} {251, 505}

\bibitem[\protect\citeauthoryear{{M{\"u}ller}, {Janka}  \&
  {Dimmelmeier}}{{M{\"u}ller} et~al.}{2010}]{mueller_10}
{M{\"u}ller} B.,  {Janka} H.,   {Dimmelmeier} H.,  2010, \mn@doi [\apjs]
  {10.1088/0067-0049/189/1/104}, \href
  {http://adsabs.harvard.edu/abs/2010ApJS..189..104M} {189, 104}

\bibitem[\protect\citeauthoryear{{M{\"u}ller}, {Janka}  \&
  {Marek}}{{M{\"u}ller} et~al.}{2012a}]{mueller_12a}
{M{\"u}ller} B.,  {Janka} H.-T.,   {Marek} A.,  2012a, \mn@doi [\apj]
  {10.1088/0004-637X/756/1/84}, \href
  {http://adsabs.harvard.edu/abs/2012ApJ...756...84M} {756, 84}

\bibitem[\protect\citeauthoryear{{M{\"u}ller}, {Janka}  \&
  {Heger}}{{M{\"u}ller} et~al.}{2012b}]{mueller_12b}
{M{\"u}ller} B.,  {Janka} H.-T.,   {Heger} A.,  2012b, \mn@doi [\apj]
  {10.1088/0004-637X/761/1/72}, \href
  {http://adsabs.harvard.edu/abs/2012ApJ...761...72M} {761, 72}

\bibitem[\protect\citeauthoryear{{M{\"u}ller}, {Janka}  \&
  {Marek}}{{M{\"u}ller} et~al.}{2013}]{mueller_13}
{M{\"u}ller} B.,  {Janka} H.-T.,   {Marek} A.,  2013, \mn@doi [\apj]
  {10.1088/0004-637X/766/1/43}, \href
  {http://adsabs.harvard.edu/abs/2013ApJ...766...43M} {766, 43}

\bibitem[\protect\citeauthoryear{{M{\"u}ller}, {Heger}, {Liptai}  \&
  {Cameron}}{{M{\"u}ller} et~al.}{2016a}]{mueller_16a}
{M{\"u}ller} B.,  {Heger} A.,  {Liptai} D.,   {Cameron} J.~B.,  2016a, \mn@doi
  [\mnras] {10.1093/mnras/stw1083}, \href
  {http://adsabs.harvard.edu/abs/2016MNRAS.460..742M} {460, 742}

\bibitem[\protect\citeauthoryear{{M{\"u}ller}, {Viallet}, {Heger}  \&
  {Janka}}{{M{\"u}ller} et~al.}{2016b}]{mueller_16c}
{M{\"u}ller} B.,  {Viallet} M.,  {Heger} A.,   {Janka} H.-T.,  2016b, \mn@doi
  [\apj] {10.3847/1538-4357/833/1/124}, \href
  {http://adsabs.harvard.edu/abs/2016ApJ...833..124M} {833, 124}

\bibitem[\protect\citeauthoryear{{M{\"u}ller}, {Melson}, {Heger}  \&
  {Janka}}{{M{\"u}ller} et~al.}{2017}]{mueller_17}
{M{\"u}ller} B.,  {Melson} T.,  {Heger} A.,   {Janka} H.-T.,  2017, \mn@doi
  [\mnras] {10.1093/mnras/stx1962}, \href
  {http://adsabs.harvard.edu/abs/2017MNRAS.472..491M} {472, 491}

\bibitem[\protect\citeauthoryear{{M{\"u}ller} et~al.,}{{M{\"u}ller}
  et~al.}{2019}]{mueller_19}
{M{\"u}ller} B.,  et~al., 2019, \mn@doi [\mnras] {10.1093/mnras/stz216}, \href
  {http://adsabs.harvard.edu/abs/2019MNRAS.484.3307M} {484, 3307}

\bibitem[\protect\citeauthoryear{{Murphy} \& {Meakin}}{{Murphy} \&
  {Meakin}}{2011}]{murphy_11}
{Murphy} J.~W.,  {Meakin} C.,  2011, \mn@doi [\apj]
  {10.1088/0004-637X/742/2/74}, \href
  {http://adsabs.harvard.edu/abs/2011ApJ...742...74M} {742, 74}

\bibitem[\protect\citeauthoryear{{Murphy}, {Ott}  \& {Burrows}}{{Murphy}
  et~al.}{2009}]{murphy_09}
{Murphy} J.~W.,  {Ott} C.~D.,   {Burrows} A.,  2009, \mn@doi [\apj]
  {10.1088/0004-637X/707/2/1173}, \href
  {http://adsabs.harvard.edu/abs/2009ApJ...707.1173M} {707, 1173}

\bibitem[\protect\citeauthoryear{{Murphy}, {Dolence}  \& {Burrows}}{{Murphy}
  et~al.}{2013}]{murphy_12}
{Murphy} J.~W.,  {Dolence} J.~C.,   {Burrows} A.,  2013, \mn@doi [\apj]
  {10.1088/0004-637X/771/1/52}, \href
  {http://adsabs.harvard.edu/abs/2013ApJ...771...52M} {771, 52}

\bibitem[\protect\citeauthoryear{{O'Connor} \& {Couch}}{{O'Connor} \&
  {Couch}}{2018}]{oconnor_18b}
{O'Connor} E.~P.,  {Couch} S.~M.,  2018, \mn@doi [\apj]
  {10.3847/1538-4357/aadcf7}, \href
  {http://adsabs.harvard.edu/abs/2018ApJ...865...81O} {865, 81}

\bibitem[\protect\citeauthoryear{{O'Connor} \& {Ott}}{{O'Connor} \&
  {Ott}}{2010}]{oconnor_10}
{O'Connor} E.,  {Ott} C.~D.,  2010, \mn@doi [Classical and Quantum Gravity]
  {10.1088/0264-9381/27/11/114103}, \href
  {http://adsabs.harvard.edu/abs/2010CQGra..27k4103O} {27, 114103}

\bibitem[\protect\citeauthoryear{{Orlando}, {Miceli}, {Pumo}  \&
  {Bocchino}}{{Orlando} et~al.}{2016}]{orlando_16}
{Orlando} S.,  {Miceli} M.,  {Pumo} M.~L.,   {Bocchino} F.,  2016, \mn@doi
  [\apj] {10.3847/0004-637X/822/1/22}, \href
  {http://adsabs.harvard.edu/abs/2016ApJ...822...22O} {822, 22}

\bibitem[\protect\citeauthoryear{{Pejcha} \& {Thompson}}{{Pejcha} \&
  {Thompson}}{2015}]{pejcha_15a}
{Pejcha} O.,  {Thompson} T.~A.,  2015, \mn@doi [\apj]
  {10.1088/0004-637X/801/2/90}, \href
  {http://adsabs.harvard.edu/abs/2015ApJ...801...90P} {801, 90}

\bibitem[\protect\citeauthoryear{{Perego}, {Hempel}, {Fr{\"o}hlich}, {Ebinger},
  {Eichler}, {Casanova}, {Liebend{\"o}rfer}  \& {Thielemann}}{{Perego}
  et~al.}{2015}]{perego_15}
{Perego} A.,  {Hempel} M.,  {Fr{\"o}hlich} C.,  {Ebinger} K.,  {Eichler} M.,
  {Casanova} J.,  {Liebend{\"o}rfer} M.,   {Thielemann} F.-K.,  2015, \mn@doi
  [\apj] {10.1088/0004-637X/806/2/275}, \href
  {http://adsabs.harvard.edu/abs/2015ApJ...806..275P} {806, 275}

\bibitem[\protect\citeauthoryear{{Shu}}{{Shu}}{1992}]{shu}
{Shu} F.~H.,  1992, {Physics of Astrophysics, Vol. II}.
University Science Books

\bibitem[\protect\citeauthoryear{{Spiegel}}{{Spiegel}}{1963}]{spiegel_63}
{Spiegel} E.~A.,  1963, \mn@doi [\apj] {10.1086/147628}, \href
  {http://adsabs.harvard.edu/abs/1963ApJ...138..216S} {138, 216}

\bibitem[\protect\citeauthoryear{{Stellingwerf}}{{Stellingwerf}}{1982}]{stellingwerf_82}
{Stellingwerf} R.~F.,  1982, \mn@doi [\apj] {10.1086/160425}, \href
  {http://adsabs.harvard.edu/abs/1982ApJ...262..330S} {262, 330}

\bibitem[\protect\citeauthoryear{{Sukhbold}, {Ertl}, {Woosley}, {Brown}  \&
  {Janka}}{{Sukhbold} et~al.}{2016}]{sukhbold_16}
{Sukhbold} T.,  {Ertl} T.,  {Woosley} S.~E.,  {Brown} J.~M.,   {Janka} H.-T.,
  2016, \mn@doi [\apj] {10.3847/0004-637X/821/1/38}, \href
  {http://adsabs.harvard.edu/abs/2016ApJ...821...38S} {821, 38}

\bibitem[\protect\citeauthoryear{{Ugliano}, {Janka}, {Marek}  \&
  {Arcones}}{{Ugliano} et~al.}{2012}]{ugliano_12}
{Ugliano} M.,  {Janka} H.-T.,  {Marek} A.,   {Arcones} A.,  2012, \mn@doi
  [\apj] {10.1088/0004-637X/757/1/69}, \href
  {http://adsabs.harvard.edu/abs/2012ApJ...757...69U} {757, 69}

\bibitem[\protect\citeauthoryear{{Unno}}{{Unno}}{1967}]{unno_67}
{Unno} W.,  1967, \pasj, \href
  {http://adsabs.harvard.edu/abs/1967PASJ...19..140U} {19, 140}

\bibitem[\protect\citeauthoryear{{Vartanyan}, {Burrows}, {Radice}, {Skinner}
  \& {Dolence}}{{Vartanyan} et~al.}{2019}]{vartanyan_19}
{Vartanyan} D.,  {Burrows} A.,  {Radice} D.,  {Skinner} M.~A.,   {Dolence} J.,
  2019, \mn@doi [\mnras] {10.1093/mnras/sty2585}, \href
  {http://adsabs.harvard.edu/abs/2019MNRAS.482..351V} {482, 351}

\bibitem[\protect\citeauthoryear{{Wilson} \& {Mayle}}{{Wilson} \&
  {Mayle}}{1988}]{wilson_88}
{Wilson} J.~R.,  {Mayle} R.~W.,  1988, \mn@doi [\physrep]
  {10.1016/0370-1573(88)90036-1}, \href
  {http://adsabs.harvard.edu/abs/1988PhR...163...63W} {163, 63}

\bibitem[\protect\citeauthoryear{{Woosley} \& {Heger}}{{Woosley} \&
  {Heger}}{2007}]{woosley_07}
{Woosley} S.~E.,  {Heger} A.,  2007, \mn@doi [\physrep]
  {10.1016/j.physrep.2007.02.009}, \href
  {http://adsabs.harvard.edu/abs/2007PhR...442..269W} {442, 269}

\bibitem[\protect\citeauthoryear{{Wuchterl} \& {Feuchtinger}}{{Wuchterl} \&
  {Feuchtinger}}{1998}]{wuchterl_98}
{Wuchterl} G.,  {Feuchtinger} M.~U.,  1998, \aap, \href
  {http://adsabs.harvard.edu/abs/1998A%26A...340..419W} {340, 419}

\bibitem[\protect\citeauthoryear{{Zhou}}{{Zhou}}{2017}]{zhou_17a}
{Zhou} Y.,  2017, \mn@doi [\physrep] {10.1016/j.physrep.2017.07.005}, \href
  {http://adsabs.harvard.edu/abs/2017PhR...720....1Z} {720, 1}

\makeatother
\end{thebibliography}

%\onecolumn
\appendix
\section{Favre Decomposition of the Equations of Hydrodynamics}
\label{sec:favre}
For compressible turbulent flows, it is often advantageous to
use mass-weighted (Favre) averages for the velocity $\mathbf{v}$,
specific internal energy $\epsilon$, and mass fractions $X_i$ when decomposing
the flow into an averaged background state and turbulent
fluctuations \citep[e.g.][]{favre_65}.
Compared to
the usual Reynolds decomposition, the use of Favre
averages significantly simplifies the resulting
equations as it largely eliminates correlations with
the density fluctuations. For reference, we here
outline this procedure, and derive the Favre-averaged
continuity, momentum, and energy equation, as well
as the equation for the turbulent kinetic energy.
We assume that averages are taken over spherical shells,
and that the gravitational field is spherically symmetric.

Favre decomposition uses the same volume-weighted
averages for the density $\rho$ and the pressure $P$, which we denote with
carets or angled-brackets as
in Section~\ref{sec:background},
\begin{eqnarray}
\hat{\rho} &=&\langle \rho \rangle\\
\hat{P} &=&\langle P \rangle.
\end{eqnarray}
For any mass-specific quantity $Y$
(e.g.\ ${v_i}$, $\epsilon$, $X_i$), one instead
uses the Favre average, which we denote by a tilde,
\begin{equation}
\tilde{Y}
=\frac{\langle \rho Y\rangle}{\hat{\rho}}.
\end{equation}
and which is used to define the fluctuations $Y''$ (denoted by a double prime
to distinguish them from fluctuations around the Reynolds
average) as
\begin{equation}
    Y''=Y-\tilde{Y}.
\end{equation}

\subsection{Continuity Equation}
Averaging the continuity equation using these definitions
leads to
\begin{align}
\frac{\pd \langle \hat{\rho}+\rho'\rangle }{\pd t}
+
\nabla \cdot
\left\langle
(\hat{\rho}+\rho')
(\tilde{\mathbf{v}}+\mathbf{v}'')
\right\rangle &=0
\\
\frac{\pd \hat{\rho} }{\pd t}
+
\nabla \cdot
(\hat{\rho}\tilde{\mathbf{v}}
+\underbrace{\langle
\rho' \tilde{\mathbf{v}}
\rangle}_{=0}
+\underbrace{\langle
\rho \mathbf{v}''
\rangle}_{=0}
) &=0,
\\
\frac{\pd \hat{\rho} }{\pd t}
+
\nabla \cdot
\hat{\rho}\tilde{\mathbf{v}}
&=0.
\end{align}
Here we have exploited the general relations
\begin{gather}
    \langle \rho Y''\rangle=
    \langle (\hat{\rho}+\rho') Y''\rangle=0,\\
    \langle \rho' \tilde{Y}\rangle=0.
\end{gather}

\subsection{Momentum Equation}
The Favre-averaged momentum equation can be obtained
in a similarly straightforward manner. Expanding
in terms of averaged and fluctuating quantities,
we obtain
\begin{multline}
\frac{\pd }{\pd t}
\langle (\hat\rho+\rho') (\tilde{\mathbf{v}}+\mathbf{v}'')\rangle
+
\nabla \cdot
\langle (\hat\rho+\rho') (\tilde{\mathbf{v}}+\mathbf{v}'')
\ot
 (\tilde{\mathbf{v}}+\mathbf{v}'')\rangle
\\
+
\nabla
\langle \hat P+P' \rangle
=
-\langle \hat{\rho} +\rho'\rangle  \mathbf{g},
\end{multline}
which can be simplified to
\begin{equation}
\frac{\pd \hat{\rho} \tilde{\mathbf{v}}}{\pd t} 
+
\nabla \cdot
(
\hat{\rho} \tilde{\mathbf{v}}\ot \tilde{\mathbf{v}}
+\underbrace{
\langle 
\rho'\tilde{\mathbf{v}}\ot \tilde{\mathbf{v}}+
\rho \mathbf{v}'' \ot \tilde{\mathbf{v}}+
\rho \tilde{\mathbf{v}}  \ot \mathbf{v}''
\rangle
}_{=0}
+
\langle\rho  \mathbf{v}''\ot \mathbf{v}'' \rangle)
+
\nabla \hat{P}
=
\hat{\rho} \mathbf{g},
\end{equation}
and,
\begin{equation}
\frac{\pd \hat{\rho} \tilde{\mathbf{v}}}{\pd t} 
+
\nabla \cdot
(
\hat{\rho} \tilde{\mathbf{v}} \ot \tilde{\mathbf{v}}
+
\hat{\rho} \widetilde{\mathbf{v}''\ot \mathbf{v}''} \rangle
)+
\nabla \hat{P}
=
-\hat{\rho} \mathbf{g}.
\end{equation}
Note that we have omitted the outer product
symbol in dyadic
products for the sake of brevity and write, e.g., 
$\mathbf{v}\mathbf{v}$ instead of
$\mathbf{v}\otimes \mathbf{v}$.
If we introduce the Favre-averaged turbulent stress
tensor $\tilde{\mathbf{R}}=\widetilde{\mathbf{v} \mathbf{v}}$,
we can write this in essentially the same form
as Equation~(\ref{eq:gl2}),
\begin{equation}
\frac{\pd \hat{\rho} \mathbf{v}}{\pd t} 
+
\nabla \cdot \hat{\rho} \tilde{\mathbf{v}} \ot \tilde{\mathbf{v}}
+
\nabla \cdot \hat{\rho} \tilde{\mathbf{R}}
+
\nabla \hat{P}
=
\hat{\rho} \mathbf{g}.
\label{eq:favre_momentum}
\end{equation}
The only difference is that the term
$\langle \rho \mathbf{v}\rangle$ now factorises
as
$\langle\rho v_r \rangle = \hat{\rho} \tilde{v}_r$.
Thus, it is actually consistent to compute the
fluid velocity $\tilde{v}_r$ as the ratio of the evolved momentum
density and mass density; one need not assume that
the turbulent mass flux vanishes in order to obtain
a simple and easily tractable form of the left-hand side
of the momentum equation.

\subsection{Energy Equation}
The Favre-averaged total energy equation is
\begin{multline}
\frac{\pd }{\pd t}
\Big\langle
(\hat{\rho}+\rho') (\tilde{\epsilon}+ \epsilon'')
+\frac{1}{2}
(\hat{\rho}+\rho') (\tilde{\mathbf{v}}+ \mathbf{v}'')\cdot
(\tilde{\mathbf{v}}+ \mathbf{v}'')
\Big \rangle
\\
+
\nabla \cdot
\Big\langle
(\hat{\rho}+\rho') (\tilde{\epsilon}+ \epsilon'')
(\tilde{\mathbf{v}}+ \mathbf{v}'')
+
(\hat{P}+P')
(\tilde{\mathbf{v}}+ \mathbf{v}'')
\\
+\frac{1}{2}
(\hat{\rho}+\rho')
(\tilde{\mathbf{v}}+ \mathbf{v}'')
(\tilde{\mathbf{v}}+ \mathbf{v}'')\cdot
(\tilde{\mathbf{v}}+ \mathbf{v}'')
\Big \rangle
=
\langle
(\hat{\rho}+\rho')
(\tilde{\mathbf{v}}+ \mathbf{v}'')
\rangle
\cdot \mathbf{g}.
\end{multline}
Expanding the various products and retaining only
non-vanishing terms (containing either
averaged quantities only, or at least two fluctuation
terms), we obtain
\begin{multline}
\frac{\pd }{\pd t}
\left(
\hat{\rho}\tilde{\epsilon}+
+
\hat{\rho} \frac{|\tilde{\mathbf{v}}|^2}{2}
+
\hat{\rho} \frac{\langle|\tilde{\mathbf{v}}''|^2\rangle}{2}
\right)
+
\nabla \cdot
\left[
(\hat{\rho}\tilde{\epsilon}+\hat{P})
\tilde{\mathbf{v}}
+
\langle \rho  \epsilon'' \mathbf{v}''\rangle
+
\langle P' \mathbf{v}''\rangle
\right]
\\
+\nabla\cdot
\left\langle
\hat{\rho} \frac{|\tilde{\mathbf{v}}|^2}{2} \tilde{\mathbf{v}}+
\rho\mathbf{v}'' (\tilde{\mathbf{v}}\cdot \mathbf{v}'')
+
\rho\tilde{\mathbf{v}} \frac{|\mathbf{v}''|^2}{2}
+
\rho\mathbf{v}''\frac{|\mathbf{v}''|^2}{2}
\right \rangle
=
\hat\rho
\tilde{\mathbf{v}}
\cdot \mathbf{g}.
\end{multline}
If we adopt the notation
$\varpi=\langle |\mathbf{v}''|^2\rangle/2$
for the turbulent kinetic energy per unit mass
and note that
$\mathbf{v''}(\tilde{\mathbf{v}} \cdot \mathbf{v}'')=
(\mathbf{v''} \mathbf{v''})\cdot \tilde{\mathbf{v}}$,
we obtain an equation that already closely resembles
Equation~(\ref{eq:kuhfuss_energy}),
\begin{multline}
\frac{\pd }{\pd t}
\left(
\hat{\rho}\tilde{\epsilon}+
+
\hat{\rho} \frac{|\tilde{\mathbf{v}}|^2}{2}
+
\hat{\rho} \varpi
\right)
+
\nabla \cdot
\Big [
(\hat{\rho}\tilde{\epsilon}+
\hat{\rho} \frac{|\tilde{\mathbf{v}}|^2}{2}
+\hat{P})
\tilde{\mathbf{v}}
+
\hat{\rho} \widetilde {\epsilon'' \mathbf{v}''}
\\
+
\langle P' \mathbf{v}''\rangle
+
\hat{\rho} \tilde{\mathbf{R}} \cdot \tilde{\mathbf{v}}
+
\hat{\rho} \tilde{\mathbf{v}} \varpi
+
\langle \rho \mathbf{v}'' \frac{|\mathbf{v}''|^2}{2}\rangle
\Big ]
=
\hat\rho
\tilde{\mathbf{v}}
\cdot \mathbf{g}.
\label{eq:favre_energy}
\end{multline}
This is a conservation equation for the sum of the
bulk internal and kinetic energy and the turbulent
kinetic energy. With one exception, 
all of the flux terms are included in the model
of \citet{kuhfuss_86}, namely
the standard total energy flux
$(\hat{\rho}\tilde{\epsilon}+\hat\rho |\tilde{\mathbf{v}}|^2/2+\hat{P}) \tilde{\mathbf{v}}$
for the spherically averaged flow,
the convective energy flux
$F_\mathrm{conv}=\hat{\rho}
\widetilde{\epsilon'' \mathbf{v}''}$,
the term $\hat{\rho}\tilde{\mathbf{R}}\cdot \tilde{\mathbf{v}}$
for work done by Reynolds stresses\footnote{Remember that this term is split into a term $P_t \mathbf{v}$ containing the turbulent
pressure $P_t$ and a viscous energy flux term in the
Kuhfuss model.},
the advective flux $\hat{\rho}\tilde{\mathbf{v}} \varpi$ of turbulent kinetic energy,
and the turbulent flux
$F_\varpi=\langle \rho \mathbf{v}'' |\mathbf{v}''|^2/2\rangle$ of turbulent kinetic energy. There is no energy source term
from buoyant driving on the right-hand side. The only term
that is explicitly discarded in the model of 
\citet{kuhfuss_86} is the acoustic energy flux $\langle P'\mathbf{v}''\rangle$ because  of the assumption of instant horizontal
pressure equilibration ($P'=0$).

\subsection{Turbulent Kinetic Energy Equation}
In order to derive the Favre-averaged equation,
we first consider the equation for the total kinetic energy
\citep{shu}:
\begin{equation}
\frac{\pd \rho |\mathbf{v}|^2}{\pd t}
        +\nabla \cdot 
    \left(\rho \mathbf{v} \frac{|\mathbf{v}|^2}{2}\right)
    +\mathbf{v}\cdot \nabla P=
    \rho \mathbf{v} \mathbf{g}.
\end{equation}
Performing a Favre-decomposition of this equation
 requires very much the same steps that
led to Equation~(\ref{eq:favre_energy}) and yields,
\begin{multline}
\label{eq:favre_kinetic}
\frac{\pd \hat\rho (|\tilde{\mathbf{v}}|^2/2+\varpi)}{\pd t}
        +\nabla \cdot 
    \left(\hat{\rho} \tilde{\mathbf{v}} \frac{|\tilde{\mathbf{v}}|^2}{2}\right)
    +
    \langle(\tilde{\mathbf{v}}+\mathbf{v}'')\cdot \nabla (\hat{P}+P')\rangle
    \\
    +\hat{\rho} \tilde{\mathbf{R}} \cdot \tilde{\mathbf{v}}
    +\nabla\cdot (\hat{\rho} \tilde{\mathbf{v}} \varpi)
    +\langle \rho \mathbf{v}'' \frac{|\mathbf{v}''|^2}{2}\rangle
    =
    \langle (\rho+\rho') (\tilde{\mathbf{v}}+\mathbf{v}'')\rangle \cdot
    \mathbf{g}.
\end{multline}
Manipulating the Favre-averaged momentum
Equation~(\ref{eq:favre_momentum})
to obtain the time derivative of $\tilde{\mathbf{v}}$,
\begin{equation}
\frac{\pd \tilde{\mathbf{v}}}{\pd t} 
+
(\tilde{\mathbf{v}}\cdot
\nabla ) \tilde{\mathbf{v}}
+
\frac{1}{\hat{\rho}} \nabla \cdot \hat{\rho} \tilde{\mathbf{R}}
+
\frac{\nabla \hat{P}}{\hat{\rho}}
=
\mathbf{g},
\end{equation}
then allows us formulate an Equation for
the kinetic energy of the bulk flow,
\begin{equation}
\frac{\pd \rho |\tilde{\mathbf{v}}|^2}{\pd t}
        +\nabla \cdot 
    \left(\hat{\rho} \tilde{\mathbf{v}} \frac{\tilde{|\mathbf{v}|^2}}{2}\right)
    +\tilde{\mathbf{v}}\cdot \nabla \hat{P}
    +\tilde{\mathbf{v}}\cdot \nabla \cdot \hat{\rho}\tilde{\mathbf{R}}
    =
    \hat{\rho} \tilde{\mathbf{v}} \mathbf{g}.
\end{equation}
After subtracting this equation
from (\ref{eq:favre_kinetic}), we obtain
the Favre-averaged equation for the
turbulent kinetic energy $\varpi$,
\begin{equation}
\frac{\pd \hat\rho \varpi}{\pd t}
    +\langle \mathbf{v}''\cdot \nabla P'\rangle
    \\
    +(\hat{\rho} \tilde{\mathbf{R}} \cdot \nabla) \tilde{\mathbf{v}}
    +\nabla \cdot (\hat{\rho} \tilde{\mathbf{v}} \varpi)
    +\langle \rho \mathbf{v}'' \frac{|\mathbf{v}''|^2}{2}\rangle
    =
    \langle \rho'\mathbf{v}''\rangle \cdot
    \mathbf{g}.
\end{equation}
Again, the model of Kuhfuss includes all of these terms
except the for the one
containing pressure fluctuations $P'$, which are assumed to
vanish. The term $(\hat{\rho} \tilde{\mathbf{R}} \cdot \nabla) \tilde{\mathbf{v}}$ accounts both for $P \,\ud V$ work
by the turbulent pressure and for turbulent energy generation
by turbulent viscous stresses, 
$\hat{\rho} \tilde{\mathbf{v}}$ accounts for the advection
of turbulent kinetic energy, and 
$\langle \rho \mathbf{v}'' |\mathbf{v}''|^2/2\rangle$
accounts for turbulent diffusion of turbulent kinetic energy.
On the right-hand side, 
we naturally recover the term $\langle \rho'\mathbf{v}''\rangle \cdot \mathbf{g}$ for buoyant generation of turbulent
kinetic energy.

\subsection{Advection of Mass Fractions}
Favre-averaging the equation for the advection of mass fractions,
\begin{equation}
\frac{\pd \langle (\hat{\rho }+\rho') (\tilde{X}_i+X_i'')\rangle}{\pd t} 
+
\nabla \cdot \langle
(\hat{\rho}+\rho')
(\tilde{\mathbf{v}}+\mathbf{v}'')
(\tilde{X}_i+X_i'')\rangle
=0,
\end{equation}
immediately leads to
\begin{equation}
\frac{\pd \hat{\rho } \tilde{X}_i}{\pd t} 
+
\nabla  \cdot
(\hat{\rho} \tilde{\mathbf{v}} \tilde{X}_i)
+\nabla
\cdot( \rho  \widetilde{\mathbf{v}'' X_i''})
=0.
\end{equation}

\section{Mixing Entropy at Constant
Pressure for General
Equation of State}
\label{sec:mixing}
Let us consider the final thermodynamic state
of two parcels of gas of equal mass with entropy $\tilde{s}\pm\delta s$
and mole fractions
$\tilde{Y}_i \pm \delta{Y}_i$.
If the mixing occurs at constant pressure $P$,
the total enthalpy is conserved, and
hence the specific enthalpy $h$ is
the mean of the initial specific enthalpies.
Thus, the specific entropy $\bar{s}$
and the mole fractions $\bar{Y}_i$
for the mixed state obey
\begin{equation}
    h(\bar{s}, P,\bar{Y}_i)= 
    \frac{
    h(\tilde{s}+\delta s, P,\tilde{Y}_i+\delta{Y}_i)+
    h(\tilde{s}-\delta s, P,\tilde{Y}_i-\delta{Y}_i)}{2}.
\end{equation}
Since the partial mass of each species
$i$ is conserved, we have
\begin{equation}
\bar{Y}_i =\frac{(\tilde{Y}_i+\delta{Y}_i)+
(\tilde{Y}_i-\delta{Y}_i)}{2}
=\tilde{Y}_i.
\end{equation}
To obtain the specific entropy for the
mixed state, we expand around
$\tilde{s}$ and $\tilde{Y}_i$, noting
that the zeroth-order terms are
identical on both sides and that
the first-order terms vanish on the
RHS,
\begin{equation}
\label{eq:b3}
\left(\frac{\pd h}{\pd s}\right)
(\bar{s}-\tilde{s})
=
\left(\frac{\pd^2 h}{\pd s^2}\right)
\delta s^2
+
2\sum_i
\left(\frac{\pd^2 h}{\pd s \pd Y_i}\right)
\delta s\, \delta Y_i
+
\sum_{i,j}
\left(\frac{\pd^2 h}{\pd Y_i \pd Y_j}\right)
\delta Y_i\, \delta Y_j.
\end{equation}
Note that we did not denote the variables that are kept constant in the derivatives
to avoid clutter; since we are only using
the natural variables of the enthalpy
no confusion arises.
The second derivatives on the RHS can
be written as derivatives
of the temperature and chemical potentials
to obtain the mixing entropy
$\bar{s}-\tilde{s}$ as
\begin{multline}
\bar{s}-\tilde{s}
=
\frac{1}{\tilde{T}}
\Bigg[
\left(\frac{\pd T}{\pd s}\right)_{P,Y_i}
\delta s^2
+
\sum_i
\left(\frac{\pd T}{\pd Y_i}\right)_{P,s}
\delta s\, \delta Y_i
\\
+
\sum_i
\left(\frac{\pd \mu_i}{\pd s}\right)_{P,Y_i}
\delta s\, \delta Y_i
+
\sum_{i,j}
\left(\frac{\pd \mu_i}{\pd Y_j}\right)_{P,s,Y_k,
k\ne j}
\delta Y_i\, \delta Y_j
\Bigg],
\end{multline}
where we have split the middle 
term in Equation~(\ref{eq:b3}) into 
derivatives of $T$ and $\mu_i$.
This is the general expression for the
mixing entropy at constant pressure
for small $\delta s$ and $\delta Y_i$.

\end{document}